\documentclass{aa}  
\usepackage{graphicx,natbib}
\usepackage{multirow}
\usepackage{siunitx}
\usepackage{epsfig}
\usepackage{rotating}
\usepackage{amsmath,amssymb,color,latexsym,longtable,array}
\usepackage{graphicx}
\usepackage{adjustbox} 
\usepackage{lipsum} 
\usepackage{placeins}
\usepackage{txfonts}
\usepackage{hyperref}
\bibpunct{(}{)}{;}{a}{}{,}

\def\ser{S\'{e}rsic\ }
\def\sser{S\'{e}rsic}

\begin{document} 
   \title{Adaptive optics imaging of the nuclei of Seyfert~2 galaxies: An unexpectedly high incidence of nuclear stellar rings 
   }
   \titlerunning{Adaptive optics imaging of the nuclei of Seyfert~2 galaxies}

\makeatletter
\renewcommand*\aa@offprintsname{\normalfont\textsuperscript{*}\ Corresponding authors}
\makeatother
\author{
N. Mackensen\inst{1,*} \and
J. Heidt\inst{1,*} \and
F. Pozo Nu\~nez\inst{2,*}
}

\offprints{%
\raggedright
nmackensen@lsw.uni-heidelberg.de;
jheidt@lsw.uni-heidelberg.de;
francisco.pozon@gmail.com}

\idline{A\&A, 711, A169 (2026)}

\institute{
Landessternwarte, Zentrum f\"ur Astronomie der Universit\"at Heidelberg,
K\"onigstuhl 12, 69117 Heidelberg, Germany
\and
Astroinformatics, Heidelberg Institute for Theoretical Studies,
Schloss-Wolfsbrunnenweg 35, 69118 Heidelberg, Germany
}

   \date{Received 8 December 2025 / Accepted 29 May 2026}

 \abstract
   {
   The high-resolution imaging of active galactic nuclei (AGNs) on subkiloparsec scales offers an important avenue for investigating their fueling.
   Observations in the near-infrared (NIR) are especially valuable as they minimize the absorption by dust.
   However, the inferred nuclear morphology may depend on the method used to analyze the images. }
   {The aim of this study is to assess whether ground-based adaptive-optics observations of the cores of Seyfert~2 galaxies in the \textit{Ks} band provide advantages over HST observations at shorter wavelengths 
   (i.e., in the \textit{V} and \textit{H} bands). We also investigate whether a dedicated 2D analysis is preferable to a 1D approach.
   } {A sample of 18  
   Seyfert~2 galaxies was observed with the adaptive optics (AO) system in the \textit{Ks} band at the Large Binocular Telescope (LBT) and compared to archival HST \textit{V}- or \textit{H}-band images. 
   The analysis included a 2D modeling procedure via GALFIT as well as an unsharp masking technique.}
   {The results obtained in different filters are mutually consistent, indicating no clear 
   advantage when using a redder filter. 
   Using both GALFIT and unsharp-masking in tandem is preferable,
   as the two methods provide complementary strengths.
   We identified nuclear stellar rings in 8 of the 18 galaxies in our sample (44$\pm$12\%). This fraction is significantly higher than reported in previous studies and about a factor of 2 higher than what was reported in the most complete atlas of nuclear rings. The radii, size distribution, and inferred masses of the detected nuclear rings are similar to those observed in non-active galaxies. 
   Low-luminosity AGNs seem to have little (if any) impact on the formation and evolution of nuclear rings.} 
  {The high incidence of nuclear stellar rings 
  in our study is unexpected and warrants further investigation. This could be tested further using the large number of suitable archival HST images available. If confirmed, it would imply that nuclear stellar rings are considerably more common 
  than previously recognized.}


   \maketitle
 \nolinenumbers
\section{Introduction}

Active galactic nuclei (AGNs) span a wide range of observed properties, driven primarily by differences in luminosity or accretion power, orientation, radio loudness, and selection method \citep[e.g.][and references therein]{2017A&ARv..25....2P}.
Classical high-luminosity AGNs are known as quasi-stellar objects (QSOs), whereas low-luminosity AGNs are classed as Seyfert galaxies.
Orientation effects (i.e., obscuration) give rise to different subclasses such as Seyfert 1 and Seyfert~2 galaxies, while radio-loudness distinguishes, for example, radio-loud quasars from radio-quiet ones.
Nonetheless, all objects in these classes share a number of properties in common. In particular, every AGN must have a supermassive black hole (SMBH) at its center
(\citealt{1995ARA&A..33..581K,2013ARA&A..51..511K}). 
The accretion of material onto the SMBH is the only way to generate the enormous output observed in these sources across the entire electromagnetic spectrum. 
Every galaxy above a certain mass threshold (thereby hosting a SMBH at its center) will most likely undergo at least one (and possibly several) activity cycles during which it can be temporarily observed as an AGN
(\citealt{2015MNRAS.451.2517S,2020MNRAS.499..334S,2026MNRAS.tmp..443H}). 
The natural questions arising from these observational properties address the processes triggering AGN activity and how it is sustained. 
Investigating these fueling processes is therefore essential, as AGN activity cannot be initiated or maintained without a steady influx of matter onto the central engine.

A number of fueling mechanisms have been proposed, depending on the luminosity and corresponding accretion power required.
The most violent mechanism is galaxy interaction or merging, which can destabilize large amounts of gas and drive inflows toward the central regions \citep{2014MNRAS.437.3373N,2014MNRAS.445..823H,2014A&A...569A..37M}.
Alternatively, chaotic cold accretion may take place, in which hot interstellar gas cools through processes such as radiative losses, collisions, and adiabatic expansion induced by interactions with other gas streams \citep{2013MNRAS.432.3401G}. 
Subsequently, the cooled gas fragments descend into the central region, where they undergo chaotic aggregation \citep{2017ApJ...837..149G}.
For more evolved hosts of low-luminosity AGNs, secular processes may become more important. 
These may include instabilities in the disk, such as bars, spiral arms, or massive clumps 
\citep{2014MNRAS.445..823H}.

In particular, galactic bars have long been proposed as
a mechanism for driving gas toward the nucleus and thereby triggering nuclear activity 
\citep{1989Natur.338...45S}, since they can induce inward gas migration through torque transfer 
and the associated removal of angular momentum. 
However, in this case the gas could stall at the inner Lindblad resonance (ILR, \citealt{1964ApNr....9..103L}) and form a ring-like structure that is often referred to as a nuclear ring (\citealt{1986ApJS...61..609B}).
All of the processes mentioned above can transport gas close to the center, but not all the way into the innermost region.
Various mechanisms have been proposed to funnel the gas through the last few hundred parsecs into the very center, including dynamical instabilities such as bars within bars, nuclear spirals, more complex nuclear structures
and/or dynamical friction (e.g., \citealt{2010MNRAS.407.1529H,2017ApJ...841L...4K}). 
\cite{2019NatAs...3...48S} offered a detailed discussion of the various fueling routes as a function of spatial scale and their associated timescales.

Optical and near-infrared HST images of the cores of Seyfert~2 galaxies have been studied, for example, by
\cite{1998ApJS..117...25M}, 
\cite{2001ApJ...562..139M}, 
\cite{2002ApJ...567...97L}, 
\cite{2002ApJ...569..624P}, 
\cite{2003ApJ...589..774M}, 
\cite{2004ApJ...616..707H}, 
\cite{2007AJ....134..648M}, 
\cite{2007ApJ...655..718S}, 
\cite{2008A&A...478..403C} 
and \cite{2010MNRAS.402.2462C}.
Depending on the scientific goal, a variety of techniques have been used, including one-dimensional (1D) analysis of surface brightness profiles based on isophotal fitting, structure-map creation, unsharp 
masking, and two-dimensional (2D) modeling. 
The results consistently showed that the vast majority of Seyfert~2 galaxies host dusty circumnuclear regions, indicating their importance for AGN fueling. 
Inner bars are less common, as are stellar rings, with the latter occurring in up to 20\% of disk galaxies in the S0--Sd range \citep{2010MNRAS.402.2462C}. 

Naturally, we would expect morphological studies of Seyfert~2 nuclei aimed at tracing the stellar component to be carried out in the \textit{K} band, where extinction along the line of sight is minimized.
Based on a compilation from the literature, \cite{2012A&A...544A.129V} estimated typical extinctions of \(A_\textit{V} = 2.5\,\mathrm{mag}\) in Seyfert~2 nuclei, with values reaching up to \(9.7\,\mathrm{mag}\) \citep{2009ApJ...698.1852B}. 
Assuming a standard Milky Way extinction law, this translates into \(A_\textit{H} - A_\textit{K} = 0.2\,\mathrm{mag}\), rising to as much as \(0.6\,\mathrm{mag}\) for the most highly obscured nuclei. 
For the galactic center, with \(A_\textit{V} \sim 30\,\mathrm{mag}\), we would expect to see an extinction difference of \(1.9\,\mathrm{mag}\), consistent with
\citet{2010A&A...511A..18S}, who derived
\(A_\textit{H} - A_\textit{K} \sim 2\,\mathrm{mag}\). 
Therefore, the gain from using the \textit{K} band instead of the \textit{H} band (the latter being the reddest band commonly used with HST) could be significant. 
However, the HST is a warm telescope, and as a result the background seen by its NIR-cameras in the \textit{K} band is about two orders of magnitude higher than in the \textit{H} band.
Consequently, there is little advantage in observing in the \textit{K} band with HST. Resolving nuclear structures in the \textit{K} band therefore requires adaptive-optics-assisted observations from the ground with 8-meter class telescopes.
Such observations are available, but they are limited to individual objects observed with IFUs (e.g., SINFONI/VLT and GEMINI/NIFS),
as detailed in
\cite{2010MNRAS.404..166R}, 
\cite{2010ApJ...713..469R},
\cite{2011MNRAS.416..493R}, 
\cite{2017MNRAS.469.3286D}, 
\cite{2017MNRAS.470..992R}, 
\cite{2019MNRAS.482.4437D},
and \cite{2019MNRAS.490.5860S}. 
These observations enable studies of not only the morphology, but also of the spatially resolved stellar populations and their kinematics in these systems.  Surprisingly, many sources show evidence of nuclear stellar rings, as indicated by drops, often called \(\sigma\)-drops, in their 2D stellar velocity dispersion field dominated by intermediate-age stars
\citep[e.g.][]{2010ApJ...713..469R,2011MNRAS.416..493R,2017MNRAS.470..992R,2019MNRAS.490.5860S}. 
Such ages are consistent with a scenario in which the origin of the rings is a past event which triggered an inflow of gas and formed stars which still keep the colder kinematics of the gas from which they have formed.

In this paper, we focus on low-luminosity AGNs, whose accretion is thought to be driven primarily by secular processes. We examine the cores of 
nearby Seyfert~2 galaxies in detail. These galaxies are particularly well suited for this purpose for two reasons.
First of all, because they are nearby, they allow us to study fueling processes on scales of tens to hundreds of parsecs. 
Secondly, because their nuclei are largely obscured by the torus, they are less dominated by direct nuclear emission than their high-luminosity counterparts. 
This minimizes the impact of the nucleus on observations of the surrounding environment on subkiloparsec scales.

We present and analyze adaptive-optics-aided, high-spatial resolution near-infrared \textit{Ks}-band images of the centers of a 
sample of 18 nearby (\(z < 0.03\))  Seyfert~2 galaxies observed with the Large Binocular Telescope (LBT). 

Our analysis has two main goals. First, we sought to investigate whether there is indeed any advantage in using the \textit{K} band instead of the \textit{H} band for this type of study.
Since the NIR background at the LBT is comparable in the \textit{H} and \textit{K} bands, any gain would be expected to come from the lower extinction in the \textit{K} band.
Our LBT images were compared to archival HST \textit{H} band (or, if unavailable, \textit{V} band) images.
We used a 2D decomposition procedure (GALFIT; \citealt{2010AJ....139.2097P}) to remove radially symmetric bulge- and disk-like components.
This allowed us to reveal, inspect, and fit the underlying (stellar) nuclear structures beneath the dominant large-scale component, as was done for NGC~2273 \citep{2023AN....34430094S}. 

Secondly, we also created unsharp-masked images \citep{2002ApJ...564..234E} to assess whether the combined use of both methods provides any advantage over previous studies, which mostly relied on a single technique, such as 1D surface-brightness profiles, structure maps, unsharp masking, or 2D modeling. 

This paper is organized as follows. In Section 2, we describe the sample, the observations,
and the data reduction, followed by the details of the fitting procedure in Section 3.
In Section 4, we summarize the results and in Section 5, we discuss the outcome, with particular focus on the unexpectedly large number of Seyfert~2 galaxies 
showing evidence of a nuclear stellar ring. In Section 6, we present the conclusions.
The images, the fits, and the corresponding best-fit-subtracted images for all sources are presented in the Appendix, where we also describe the properties of each source and the outcome of the corresponding fits individually.
A flat lambda cold dark matter (\(\Lambda\)CDM) cosmology with $H_0 = 70\,\mathrm{km\,s^{-1}\,Mpc^{-1}}$ and
$\Omega_{\mathrm{m}} = 0.3$ is assumed throughout. 

\section{Sample, observations, and data reduction}
\subsection{The sample}
We aimed to obtain a sample of about 20 Seyfert~2 galaxies to gain access to a statistically useful number of observed sources. 
The targets were selected from the QSO catalog of \cite{2010A&A...518A..10V}, imposing a redshift limit of \(z < 0.03\) to maximize spatial resolution and a brightness
limit of $m_\textit{V} < 15.0$ to ensure a sufficiently bright reference source for the
AO correction (the central two arcsec of the nucleus served as the reference source).
\begin{table*}[t]
\caption{Properties of the sample galaxies. }
\label{sample}
\centering
\resizebox{\textwidth}{!}{%
\begin{tabular}{l|crlrcccrccc}
\hline
\hline
\rule{0pt}{3ex}
Target & z & $T$ & RC3 type & $m_{\mathit{K}}$ & \textit{i}    & AGN-type & $M_{\mathrm{BH}}$    &  Scale  & Date & $T$     &   $N_{\mathrm{images}}$\\
       &   & &             & [mag] & [deg] &          & [\(\log(M_\odot)\)] &  [pc/"] &      & [sec] & [LUCI1/2]\\ 
\hline
\rule{0pt}{3ex}
MRK~461  & 0.0163 & 2 & S & 10.73  & (43.7)/43.6 & Sy2 & 7.59 & 330 & May 8 2023& 960 & -/16\\
NGC~2985 & 0.0044 & 2 & (R')SA(rs)ab &  7.36 &  (37.9)  & Sy1.9 &  8.2  & 91   &  Mar 15 2022 & 960   &  16/-  \\
NGC~3185 & 0.0041 & 1 & (R)SB(r)a   & 9.15  &  (60.1)/48.0 & Sy2   & 6.07/6.9   & 85   & May 8 2023   & 960   &  -/16 \\
NGC~3254 & 0.0045 & 4 &SA(s)bc     & 8.80  & (64.4) & Sy2  & 7.2 & 93 & May 12 2023 & 1920 & 16/16 \\
NGC~3486 & 0.0023 & 5 &SAB(r)c     & 8.00 &  (46.0)    & Sy2  & 6.14/6.5 & 48 & May 12 2023 & 1860 & 16/15\\
NGC~3607 & 0.0031 &-2 &SA0$^\wedge$0(s)? & 6.99 & (35.0)     & Sy2 & 8.08 & 64 & May 8 2023& 1980 & 16/17\\
NGC~4138 & 0.0029 &-1 &SA0$^\wedge$+(r) & 8.20 & (61.4) & Sy1.9 & - & 60 & Mar 14 2022 & 960& 16/-\\
NGC~4725 & 0.0040 & 2 &SAB(r)ab pec& 6.17 & (45.4)/42.0 & Sy2 & 7.50 & 83 & Mar 14 2022 & 900& 15/-\\ 
NGC~4941 & 0.0037 & 2 &(R)SAB(r)ab? & 8.22 & (36.6)/57.0 & Sy2 & 6.90 & 76 & May 11 2023& 960 & 16/-\\
NGC~5033 & 0.0029 & 5 &SA(s)c      & 6.96 & (64.6)/64.0 & Sy1.8 & 7.30 & 60 & Mar 15 2022 & 960& 16/-\\
NGC~5273 & 0.0036 &-2 &SA0$^\wedge$0(s) & 8.87 & (58.1)/24.0 & Sy1.9 & 6.7--7.3 & 74 & Mar 15 2022 & 780& 13/-\\
NGC~5347 & 0.0079 & 2 &(R')SB(rs)ab& 9.65 & (45.3)/45.0 & Sy2   & 8.70 & 162 & May  8 2023& 840& 14/-\\
NGC~5631 & 0.0065 &-2 &SA0$^\wedge$0(s) & 8.47 & (0)/21.0 & Sy2   & -    & 134 & Mar 15 2022 & 660& 11/-\\
NGC~5695 & 0.0142 & 2 &S?          & 9.67 & (66.5)  & Sy2   & 7.70 & 290 & May  8 2023& 1680 & 10/18\\
NGC~6211 & 0.0176 &-2 &SB0$^\wedge$0(r) pec? & 9.69 & (50.4)& Sy2 & -  & 358 & May  8 2023& 780& -/13\\
NGC~7217 & 0.0032 & 2 &(R)SA(r)ab & 6.83 & (33.4)/36.0 & S3 & 7.48  & 66& Jul 1 2023& 960& -/16\\
NGC~7466 & 0.0250 & 3 &Sb         & 10.49 & (73.0) & H\,{\sc ii}   & -     & 504 & Jul  1 2023& 1620& -/27\\
NGC~7674 & 0.0290 & 4 &SA(r)bc pec & 9.79 & (26.7)/30.0 & Sy2 & - & 581 & Jul 3 2023& 720& -/12\\
\hline
\end{tabular}%
}
\tablefoot{The redshifts are taken from NED, the Hubble \textit{T}-types from 
Hyperleda \citep{2014A&A...570A..13M}, the RC3 types from \cite{1991rc3..book.....D}, and the integrated \textit{K}-band magnitudes from 2MASS \citep{2003AJ....125..525J} or the 2MASS Extended Mission Data Release. The inclinations in brackets are estimated from Hyperleda \citep{2014A&A...570A..13M}, if measurements exist, their references are given in the Appendix \ref{appendix2}. The AGN-types are from \cite{2010A&A...518A..10V}. For NGC~3607, NGC~5033, NGC~5273, NGC~7217, and NGC~7466, the AGN classification varies in the literature, see Appendix \ref{appendix2}. We also provide the masses of the SMBHs, where available. The observational information includes the date at the start of the observing night, the integration time of the final combined image, and the number of one-minute frames from the individual LUCI instruments used to construct the final image.}
\end{table*}
\subsection{LBT observations}

The nuclei of 18 Seyfert~2 galaxies were observed during three observing runs on March 14 and 15, 2022, May 8, 11 and 12, 2023, and July 1, 2 and 3, 2023, with the Large Binocular Telescope (LBT). 
During these nights, the targets were selected randomly from the sample described above according to the available RA/Dec range, the wind direction (to avoid excessive forces on the adaptive secondaries used for the wavefront correction), and a minimum angular separation of $>$ 30\degr\ from the Moon to avoid an excessively bright background for the AO system and the telescope's acquisition and guiding unit.
As a result, the sample is not statistically complete, but it should nevertheless be largely unbiased.
The properties of the target galaxies are listed in Table~\ref{sample}.

The single conjugate adaptive optics upgrade for LBT (Soul, \citealt{2016SPIE.9909E..3VP}) supported \textit{Ks}-band images were obtained during the March 2022 and July 2023 runs with LUCI1 and during the May 2023 run in binocular mode with LUCI1 and LUCI2 (\citealt{2003SPIE.4841..962S}), respectively.
The observations were conducted with the N30 camera of the LUCI instruments, which provides a field of view (FoV) of \(30\arcsec \times 30\arcsec\) at a pixel scale of \(0\farcs015\,\mathrm{pixel}^{-1}\).
A \textit{Ks} filter was chosen for our observations because its sky background is lower than the one through the \textit{K} filter.
We employed a standard sequence for extended sources in an alternating sky-object-sky pattern. 
The sky positions were offset by at least 10\arcmin\ from the target and were checked against 2MASS and UKIDSS images to ensure (to the furthest extent possible) that they were free of stars.
Each sky frame was taken at a different position to avoid overlapping stars when constructing the final sky frames. 
The object frames were dithered by a few pixels relative to each other to maximize the common area available for analysis and to mitigate the effects of bad pixels.
In total, we obtained 16 on-target images and 10 sky images per galaxy, each with an exposure time of \(1\,\mathrm{min}\).
Each 1-min exposure consisted of the sum of six individual 10-s exposures.
This ensured that we did not suffer from saturation of the target nucleus and/or any stars in the FOV.
The weather was photometric on all nights, with an ambient seeing of 1\arcsec\ or better. 
Standard stars from \cite{1998AJ....116.2475P} were observed during the 2023 runs to determine the photometric zero point. Since no such observations were obtained during the 2022 run, we instead adopted the zero points provided on the LUCI instrument pages of the observatory.

\subsection{Data reduction}

The data reduction was carried out separately for each target using the European Southern Observatory Munich Image Data Analysis System 
(ESO-MIDAS, \citealt{1983Msngr..31...26B}). 
First, all the images were corrected for nonlinearity. 
Next, the median of all individual sky frames was determined and subtracted, and the resulting frames were median-combined to form a master sky frame.
Before combining the frames, we verified that no changes in the gradients were present. 
Then, the median of each object frame was determined and subtracted, followed by the subtraction of the master sky frame.
Finally, the science frames were divided by a flatfield constructed from a set of calibration frames obtained with the LUCI internal calibration unit. 
In this step, five lamp-on-and-off frame pairs were subtracted to remove static features, then normalized and median-combined.

For some objects, stars were present in the frames and in these cases the images were aligned using the centroids of field stars; otherwise, the alignment was based on the brightest pixel of the nucleus. 
Once aligned, the images were averaged, with bad pixels flagged and excluded from the combination.
A bad-pixel map had been created beforehand from dark frames and highly exposed flat fields.
Frames with obviously degraded image quality were not included in the stacking.
For objects with stars in the field, we typically measured a FWHM of 0\farcs1 in the combined images.
This is close to the diffraction-limit of the LBT in the \textit{K} band (0\farcs066). 

First, we combined the images from each instrument separately. 
For the data sets obtained in binocular mode in May 2023, we did not combine them blindly.  
Due to the different sensitivities of the secondary mirrors of LUCI1 and LUCI2, the image quality was expected to differ. 
This is because the AO systems of the two instruments received different numbers of photons to determine the correction.
In addition, local dome seeing affected one instrument more strongly than the other. In most, but not all, cases, the LUCI2 data provided better image quality than the LUCI1 data, as judged from the fits to the galaxy cores or to stars present in the frames. 
In the end, we combined the data from both instruments for four of the seven galaxies observed in May 2023, while for the other three galaxies we used only the data from either LUCI1 or LUCI2.
In Table~\ref{sample}, we list the total integration time and the number of images from each instrument used in constructing the final image. 

We obtained four measurements of each standard star per night. 
After the standard data reduction, we determined the zero point for each measurement individually and then averaged the results. 
To place the photometry of the science frames and the standard-star frames on the same system, the images were corrected for atmospheric extinction using data provided by the observatory and for Galactic reddening following \citet{2011ApJ...737..103S}.

\subsection{Archival HST data}

Data taken at wavelengths closest to our 
\textit{Ks}-band observations were retrieved from the archive.
For most sources, we were able to use \textit{F160W}, \textit{F187N}, or \textit{F190N} images obtained with the NICMOS cameras. 
This was not possible for MRK~461, NGC~3486, NGC~4725 and NGC~5631, for which we instead used WFPC \textit{F606W} data.
NGC~3254 was imaged with HST, but the available data were not suitable for our purposes, while NGC~7466 has not yet
been imaged with HST. 
Only the fully processed, drizzled science images from the HST archives were used. 
The dataset is summarized in Table~\ref{HSTparameter}.

\section{Analysis}
\begin{table}[ht]
    \caption{Properties of the HST images retrieved from the HST archive.}
    \label{HSTparameter}
    \centering
    \begin{tabular}{l l c r r }
        \hline
        \hline
        \rule{0pt}{3ex}
        Target& Prop-ID & Camera/Filter &$T$ [sec] & $N$ \\
        \hline
        \rule{0pt}{3ex}
        MRK~461  &  8597 & WFPC2/PC/\textit{F606W}&  560 & 2 \\
        NGC~2985 &  7330 & NIC2/\textit{F160W}&  640 & 2   \\        
        NGC~3185 & 11080 & NIC3/\textit{F190N}& 1344 &  6  \\
        NGC~3254 & -            &-      &  -  & -  \\
        NGC~3486 &  5446 & WFPC2/PC/\textit{F606W}&  160 & 2  \\
        NGC~3607 & 11219 & NIC2/\textit{F160W}& 1152  & 4  \\
        NGC~4138 & 11080 & NIC3/\textit{F187N}&  960  &  6 \\
        NGC~4725 &  8597 & WFPC2/PC/\textit{F606W}&  560 & 2  \\
        NGC~4941 &  7330 & NIC2/\textit{F160W}&  640 &2   \\
        NGC~5033 &  9360 & NIC3/\textit{F160W}&   96 & 4  \\
        NGC~5273 &  7330 & NIC2/\textit{F160W}&  640 &  2 \\
        NGC~5347 &  7867 & NIC1/\textit{F110W}& 1024 &4   \\
        NGC~5631 & 13324 & WFC3/UVIS/\textit{F814W}& 1194 & 3\\
        NGC~5695 &  7867 & NIC1/\textit{F160W}& 1024 & 4 \\
        NGC~6211 &  - & -& -  & -\\
        NGC~7217 & 11080 & NIC3/\textit{F187N}   &960 & 6 \\
        NGC~7466 &   -      & -     & - & -   \\
        NGC~7674 &  7328 & NIC1/\textit{F160W}&  768 & 3  \\

        \hline
    \end{tabular}
    \tablefoot{Listed are the proposal ID, 
    the instrument/filter combination used, the total integration time, and the number of frames used to 
    construct the final image. 
    Suitable HST data were not available for 
    NGC~3254 and NGC~7466. For NGC~6211, we were unable to obtain a reliable fit.}
\end{table}
We used two approaches to analyze the images. GALFIT \citep{2010AJ....139.2097P} was employed for a 2D decomposition of both the LBT and HST data sets. 
In both cases, we removed the underlying smooth 2D light distribution and inspected the residuals. 
For the LBT data, we additionally applied ring and bar fits to the residuals after subtraction of the smooth component. 
Since the HST data set is composed of images obtained with different cameras, filters, pixel scales, and exposure times, we used it for a quantitative comparison with the structures seen in the LBT images after removal of a smooth 2D light distribution. 
Therefore, no fits were applied to the residual HST images.

In addition, we created unsharp-masked images \citep{2002ApJ...564..234E} of the LBT data to verify that the structures detected with GALFIT are real and not artifacts of the fitting strategy.
Both approaches are described in detail below.
\subsection{GALFIT}\label{GALFIT}
GALFIT is a 2D fitting process based on a least-squares Levenberg-Marquardt algorithm, which attempts to minimize $\chi^2_{\mathrm{\nu}}$ by adjusting the free parameters step by step until a minimum is found. Here, $\chi^2_{\rm \nu}$ is defined as 

\begin{equation}
    \chi^2_{\nu} =
    \frac{1}{N_{\mathrm{dof}}}
    \sum^{n_{\mathrm{x}}}_{x=1}
    \sum^{n_{\mathrm{y}}}_{y=1}
    \frac{\left(f_{\mathrm{data}}(x,y) - f_{\mathrm{model}}(x,y)\right)^2}
    {\sigma(x,y)^2},
\end{equation}
where $n_{\mathrm{x}}$ and $n_{\mathrm{y}}$ correspond to the number of pixels in their respective directions, while $N_{\mathrm{dof}}$ refers to the number of degrees of freedom. Furthermore, $f_{\mathrm{data}}(x, y)$ denotes the individual pixel values at the positions x and y of the image provided to GALFIT for modeling. Similarly, $f_{\mathrm{model}}$ represents the pixel values of the generated model. Lastly, $\sigma(x, y)$ refers to the sigma image, which reflects the standard deviation in each pixel. 

To remove the smooth large-scale contribution from the images, we fit the general \ser profiles of the form
\begin{equation}\label{Sersicprofil}
    \Sigma(r) = \Sigma_{\mathrm{e}} \cdot
    \exp \bigg[
    -\kappa \bigg(
    \left(
    \frac{r}{r_{\mathrm{e}}}
    \right)^{\frac{1}{n}} - 1
    \bigg)
    \bigg],
\end{equation}
where $\Sigma(r)$ represents the surface brightness of the profile at a given radius, $r$, and $\Sigma_{\mathrm{e}}$ is defined as the surface brightness at the effective radius, $r_\mathrm{e}$, where half of the total luminosity is contained. To fulfill this condition, $\kappa$ cannot be a free parameter. 
The \ser index $n$ reflects the shape of the profile, with $n = 0.5$ corresponding to a Gaussian profile, $n = 1$ to an exponential profile, and $n = 4$ to a de~Vaucouleurs profile.
The radial coordinate $r$ is expressed as an ellipse by
\begin{equation}\label{Radius}
   r(x,y) =
\sqrt{
(x - x_{\mathrm{c}})^2
+
\left(
\frac{y - y_{\mathrm{c}}}{q}
\right)^2},
\end{equation}
with $x_{\mathrm{c}}$ and $y_{\mathrm{c}}$ the center coordinates, with $q$ as the axis ratio of the ellipse. 

\subsubsection{Fitting strategy: Removing the underlying smooth 2D light distribution}\label{fitstrat}
To reveal underlying structures, we began by fitting a single \ser profile to the data. However, as a single \ser profile was often insufficient to reproduce the observed light distribution, we added one (and in some cases two) additional \ser profile(s) to the fits.
The fitting procedure was iterative, similar to the approach described in \cite{2015ApJS..219....4S}; initially, a subset of parameters was fixed while the remaining parameters were allowed to vary until a convergence was reached. The resulting parameters were then used as input for the next iteration, in which additional parameters were released. Because the primary goal of the fits was to remove the 2D light distribution, rather than to derive physically meaningful bulge or disk parameters, this process was continued until a configuration was obtained in which all parameters were free to vary. 

It is tempting to expect all fitted components to share the same central coordinates and this constraint was initially imposed in the fits; however, in most cases, this was unsuccessful, as GALFIT consistently failed to converge and we found the best approach was to leave the central coordinates as free parameters for the multicomponent fits. 

\begin{figure*}
 \centering
\includegraphics[width=0.24\textwidth]{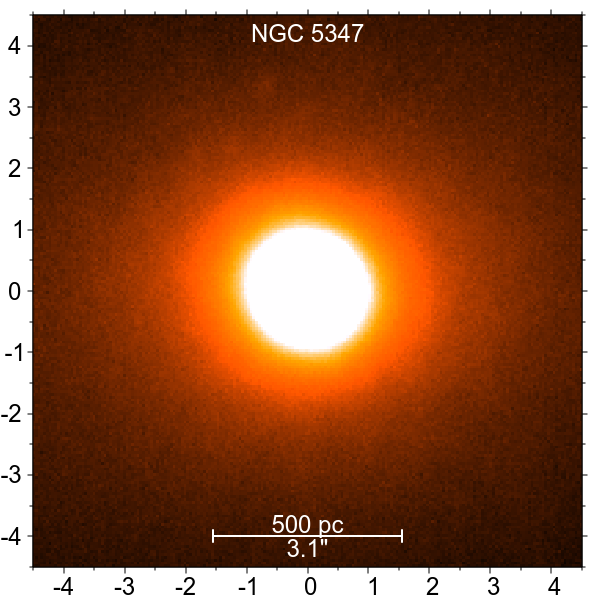}
\includegraphics[width=0.24\textwidth]{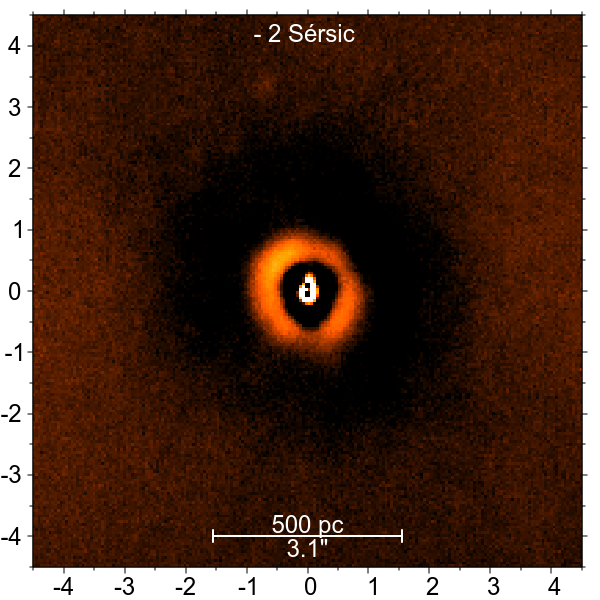}
\includegraphics[width=0.24\textwidth]{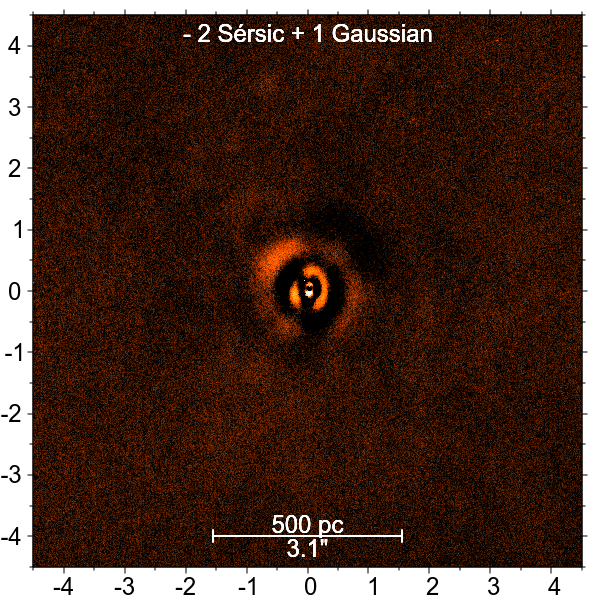}
\includegraphics[width=0.24\textwidth]{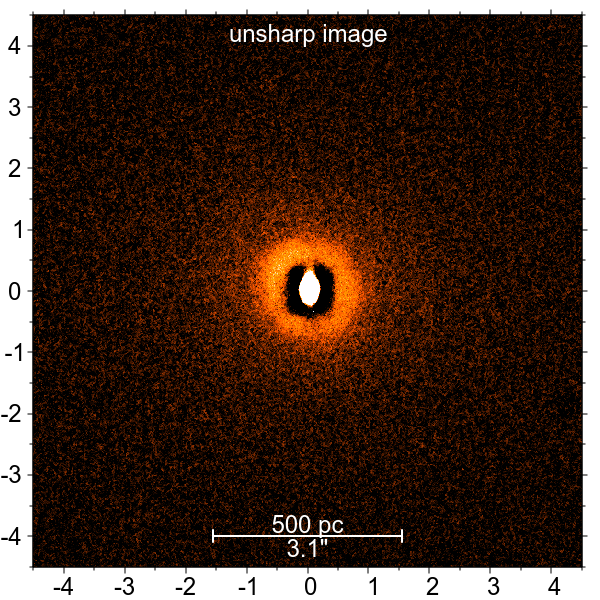}
\vspace*{.2cm}
\includegraphics[width=0.24\textwidth]{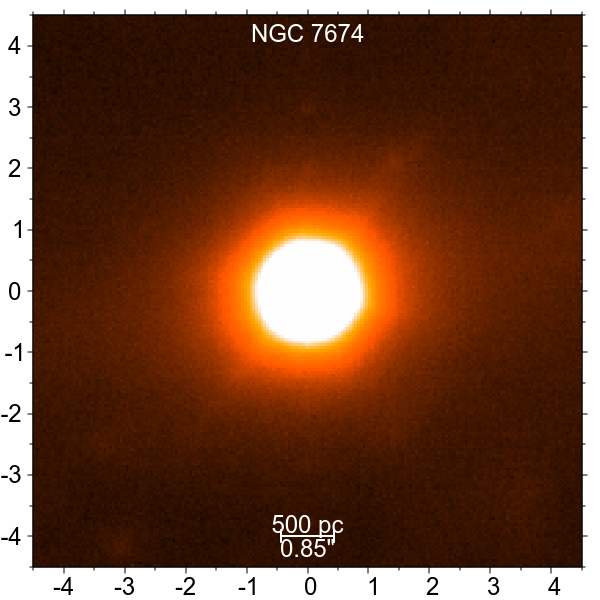}
\includegraphics[width=0.24\textwidth]{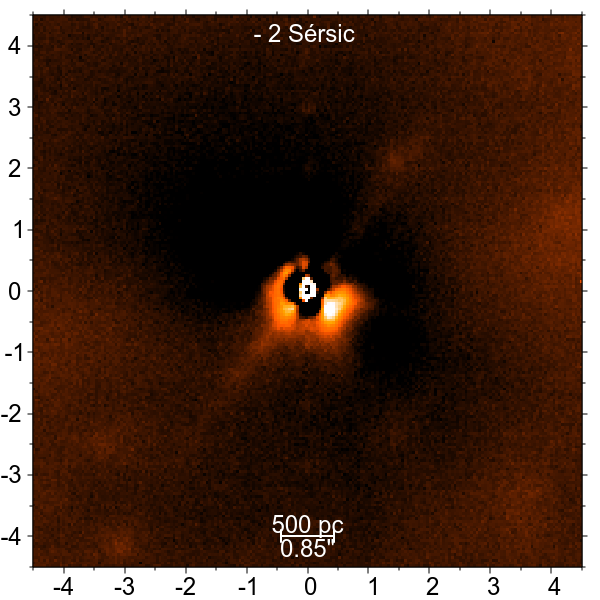}
\includegraphics[width=0.24\textwidth]{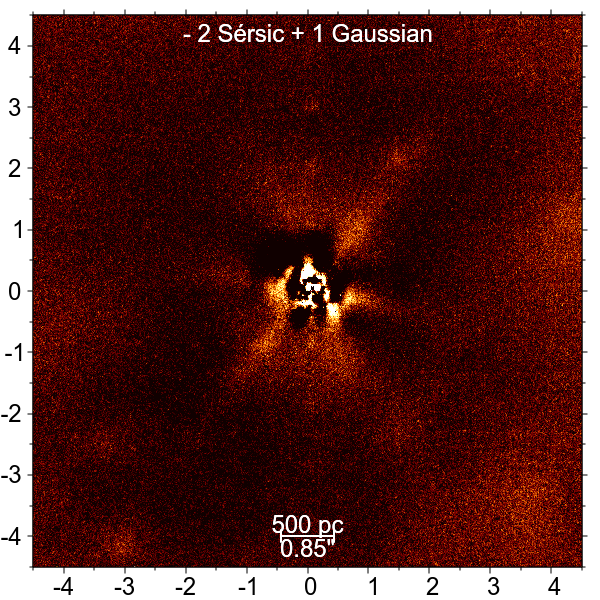}
\includegraphics[width=0.24\textwidth]{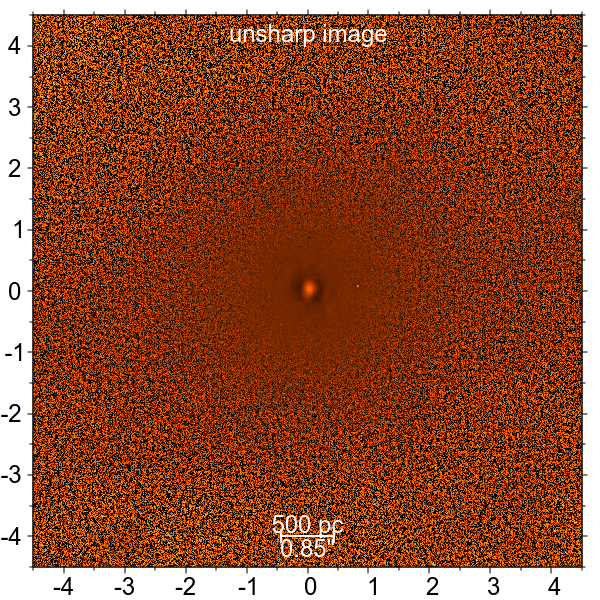}
\caption{Top: Original image of NGC~5347, together with the residual images obtained after subtraction of two \ser profiles, two \ser profiles plus one Gaussian component, and the corresponding unsharp-masked image.
Bottom: Same as above, but for NGC~7674. While a ring-like structure is present in all residual images as well as in the unsharp-masked image for NGC~5347,
a ring-like feature in NGC~7674 appears only after subtraction of two \ser profiles and disappears when a Gaussian component is included.}
\label{gausstest}
\end{figure*}

Although this approach was ultimately successful, the centers of the fitted components differed by more than \(20\,\mathrm{pixels}\) (0\farcs3) in a few galaxies, producing residuals reminiscent of unclosed ring-like structures that were not seen in the unsharp-masked images; the galaxies affected were NGC~5273, NGC~5631, and NGC~7674. 
To assess whether the ring-like residuals in these three galaxies are real or are instead artifacts of an inadequate multi-\ser fit, we first determined the FWHM of the central component in all galaxies of the sample. 
They range from \(17\) to \(55\,\mathrm{pixels}\) (0\farcs255) to (0\farcs825). 
This implies that the central component is resolved in all targets, given the diffraction limit of 0\farcs066 in our case. 
Notably, the three galaxies with residuals reminiscent of unclosed ring-like structures all had a relatively narrow central component, 
with FWHM values below \(30\,\mathrm{pixel}\). We therefore repeated the fits for these sources using one Gaussian to represent the central component 
and two \ser profiles to describe the underlying light distribution.
In all cases, the unclosed ring-like structures disappeared.
As a proof of concept, we repeated the fit for NGC~5347 in the same way, which clearly shows a nuclear ring both in the residuals from the initial fit with three \ser profiles and in the unsharp-masked image. 
In that case, the nuclear ring remained present, confirming that it is real. Figure~\ref{gausstest} illustrates this comparison for one galaxy showing a ring-like structure that is independent of whether 
a Gaussian component is included to represent the core (NGC~5347), and for one galaxy in which the ring disappeared after inclusion of such a Gaussian component (NGC~7674).

To account for the difference between the observed image and the intrinsic image of the galaxies, the models were convolved with the point spread function (PSF) during the fitting process.
Unfortunately, the PSF of the LBT data was poorly defined in our case. 
There is no direct way to determine it, since the galaxies are extended and no stars were present within the isoplanatic patch on the detector. 
Therefore, we experimented with a theoretical PSF composed of a diffraction-limited core and a seeing disk, whose parameters were estimated from the quantities used by the adaptive-optics system for the wavefront correction.  
The various solutions we attempted, including an additional seeing disk, did not improve the fits. We therefore convolved the models only with the theoretical diffraction-limited PSF during the fitting process.
The fitting strategy for the HST data was essentially identical, except that no PSF was used in the convolution.
Using a drizzled PSF from the HST pipeline also did not improve the fits.
\subsubsection{Fits to the residuals}
Both NGC~4725 and NGC~4941 exhibit complex structures characterized by a bar embedded within a ring-like feature. 
To emphasize the faint ring-like structures more clearly, we adopted the approach from \citet{2020MNRAS.491.1800B} by fitting a modified Ferrer function to model the bar component. 
GALFIT does not converge easily for NGC~4725, a problem that has previously been reported for other galaxies in \citet{2015ApJ...799...99K} and \cite{2020MNRAS.491.1800B}.
This issue is resolved by fixing the alpha parameter, which controls how sharply the profile falls off, to a value of 1.5, as recommended by \citet{2020MNRAS.491.1800B}.

Three of our objects (NGC~3185, NGC~5347, and NGC~7466) displayed fairly bright ring-like structures in their residual images. 
We attempted to model these structures to determine their physical parameters. This was done by fitting a \ser profile to the images, modified by two truncation parameters: the inner truncation, \(T_{\mathrm{i}}\), and the outer truncation, \(T_{\mathrm{o}}\).
Both \(T_{\mathrm{i}}\) and \(T_{\mathrm{o}}\) are defined by two parameters: the break radius, \(r_{\mathrm{break}}\), which marks the limit up to which the profile matches \(\SI{99}{\%}\) of the original, unaltered profile and it is not yet affected by the truncation function, and the softening length, \(\Delta r_{\mathrm{soft}}\), which indicates the radial interval over which the luminosity drops to \(\SI{1}{\%}\) from \(r_{\mathrm{break}}\).

To achieve a more realistic fit to a ring, the \ser profile can be modified using Fourier and/or bending modes. The Fourier modes modify the radius as
\begin{equation}
    r(x,y) = r_0(x,y) \cdot \bigg( 1 + \sum^{N}_{m = 1} a_{\mathrm{m}} \cos(m(\theta + \phi_{\mathrm{m}})) \bigg),
\end{equation}
where \(r_0\) denotes the original elliptical shape, which is then modified accordingly.
This modification can be applied using up to \(N\) modes, each with an amplitude \(a_{\mathrm{m}}\) that expresses the deviation of the radius relative to the unperturbed ellipse.
In addition, the description includes the position angle \(\theta\), which is already a free parameter for a standard ellipse, and the phase angle \(\phi_{\mathrm{m}}\), defined relative to the position angle of the generalized ellipse.

Bending modes introduce curvature into the model and are implemented as a modification of the \(y\)-coordinate to \(y'\) as
\begin{equation}
    y' = y + \sum^{N}_{m = 1} a_{\mathrm{m}}
    \bigg( \frac{x}{r_{\mathrm{scale}}} \bigg)^m,
\end{equation}
where \(N\) again denotes the number of modes used, and \(m\) specifies the mode order.
In addition, \(r_{\mathrm{scale}}\) denotes the scale radius of the model; in the case of the \ser profile, this corresponds to the effective radius, while \(a_{\mathrm{m}}\) represents the amplitude.

\subsection{Unsharp masking}

Unsharp masking provides an efficient way to assess whether faint substructures are embedded in the bright, diffuse body of a galaxy. 
This technique was first introduced by \cite{1979S&T....57..354M} and \cite{1985LNP...232..144S} for prints of photographic plates and was later adapted for digital images by \cite{2002ApJ...564..234E}. 
A related variant is provided by structure maps, as used by \cite{2002ApJ...569..624P}, which are based on the Richardson--Lucy image restoration technique (\citealt{1972JOSA...62...55R,1974AJ.....79..745L}).
Unsharp mask images are created by smoothing the original galaxy image and dividing the original image by the smoothed version.

Since the size of the smoothing kernel determines the characteristic width of the substructures that are enhanced, we adopted a range of smoothing kernel sizes. 
As the reference FWHM, we adopted the diffraction limit of the LBT in the \textit{K} band, with the corresponding Gaussian standard deviation $\sigma_\mathrm{PSF}$ = FWHM / (2 $\cdot \sqrt{2 \cdot \ln{2} }$). 
For the shape and orientation of the kernel, we adopted the axis ratio and position angle of the host galaxy at two half-light radii, as derived from ellipse fits to the images of the Seyfert~2 galaxies. 
We then constructed a set of Gaussian smoothing kernels with standard deviations ranging from 1 to 10 $\times$ $\sigma_\mathrm {PSF}$
and selected the one that unveiled more structures in a target-by-target basis.

Formally, the PSF in diffraction-limited observations is a combination of a diffraction-limited core and a seeing halo, with their relative contributions depending on the achieved Strehl ratio. 
However, because we used integer-pixel shifts for the alignment and co-addition of the individual frames, the PSF in the final combined image is less sharply diffraction-limited. 
Under these circumstances, a Gaussian approximation to the PSF is sufficient for our purposes.

Our unsharp-masked images provide complementary information and can be used to assess whether any of the features detected with GALFIT are real or artifacts. 
However, they become ambiguous when the substructures are large and/or diffuse, since in such cases a very large smoothing kernel would be required.
An example is NGC~3254, for which the unsharp-masked image was of limited use.
However, as discussed in Section \ref{fitstrat}, they are useful in assessing whether a detected ring-like structure is real.

\section{Results: Properties of the nuclear regions of our targets}

As expected, we find a wealth of structures close to the centers of the Seyfert~2 galaxies following the subtraction of one or more Sérsic, Ferrer and/or Gaussian components. 
Approximately two-thirds of the targets show evidence of inner bars or nuclear ring-like structures. 
Four sources are comparatively unremarkable, while two could not be fitted reliably and/or show clear signs of tidal interaction.

Surprisingly, we found ring-like structures with diameters between \(1\arcsec\) and \(10\arcsec\) in at least 8 of the 18 galaxies ($>$ 40\%); of these, 5 are new detections (NGC~4725, NGC~4941, NGC~5347, NGC~5695, and NGC~7466).  
With the exception of NGC~7466, all of these galaxies have previously been imaged with HST and analyzed, but the rings were not identified. 
Galaxies that are already known to host a nuclear ring are 
NGC~3185 \citep{2008JPhCS.131a2046C}, NGC~3607 \citep{2005AJ....129.2138L}, and NGC~7217 \citep{1917ApJ....46...24P,1989ApJS...71..433P,1995ApJ...450..593B,2004A&A...414..857C,2006AJ....131.1336S}.
In addition, three cases remain ambiguous. 
We might have detected a previously unrecognized nuclear ring at smaller radius than the already known ring in NGC~7217. Furthermore, we cannot rule out a nuclear ring-like structure in NGC~5273 and we possibly detected a partial ring in NGC~7674.
If confirmed, the fraction of galaxies with a nuclear ring-like structure in our sample would rise to 10/18 (55\%).

\begin{figure*}[h]
 \centering
\includegraphics[width=0.33\textwidth]{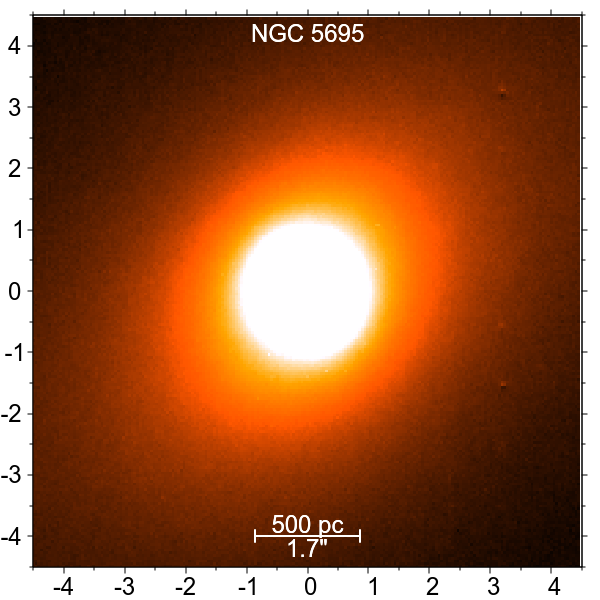}
\includegraphics[width=0.33\textwidth]{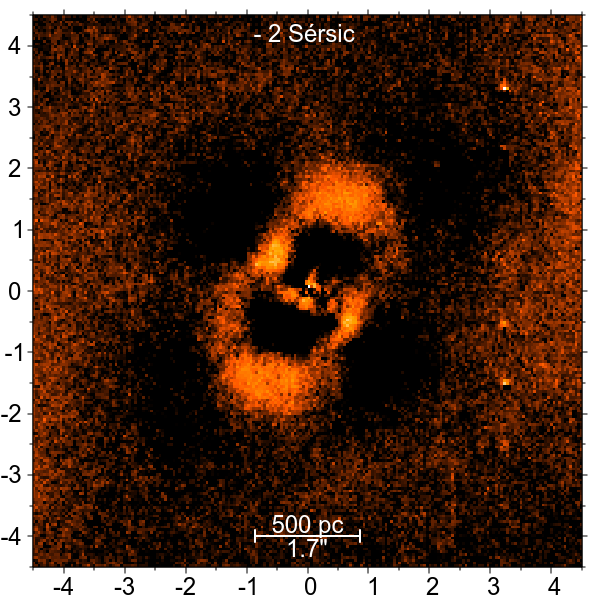}
\includegraphics[width=0.33\textwidth]{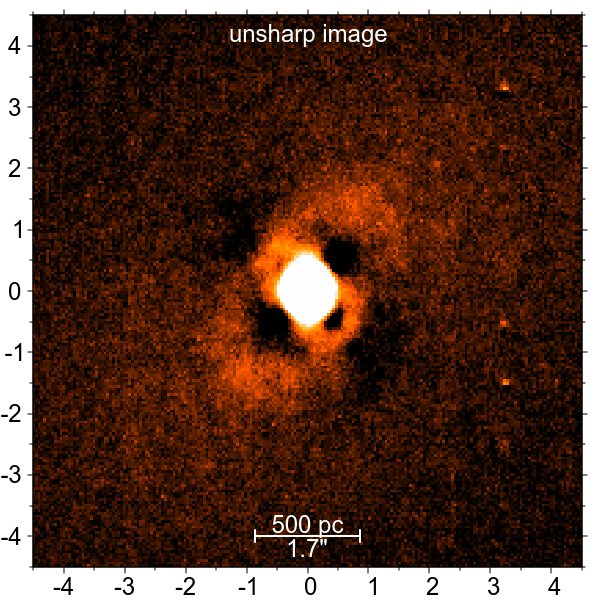}\\
\vspace*{.2cm}
\includegraphics[width=0.33\textwidth]{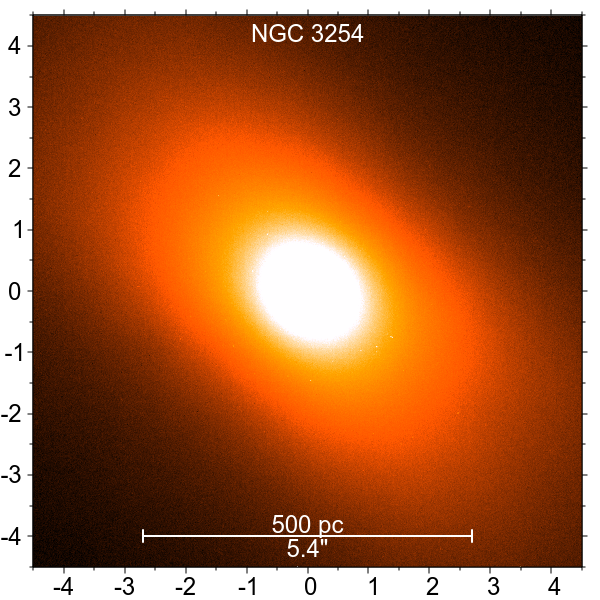}
\includegraphics[width=0.33\textwidth]{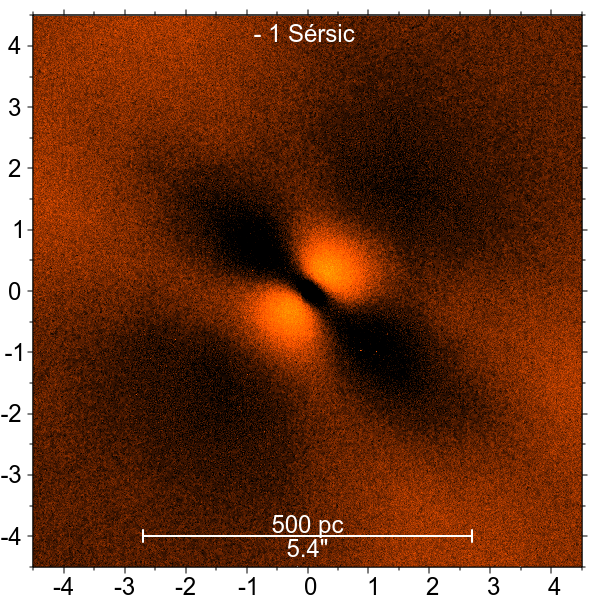}
\includegraphics[width=0.33\textwidth]{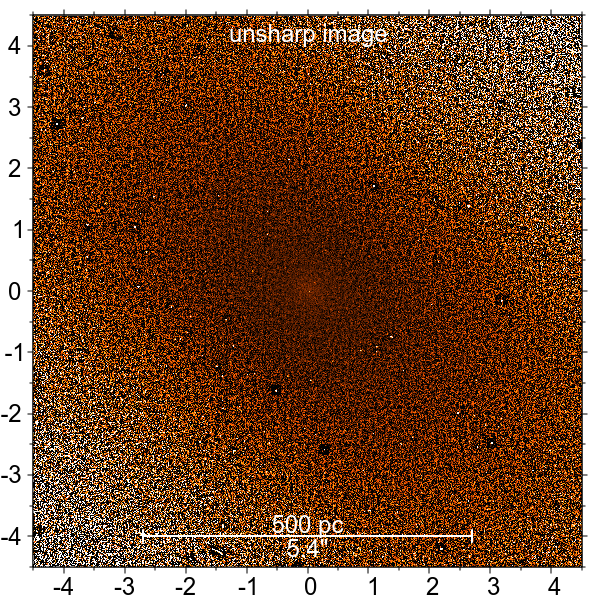}\\
\vspace*{.2cm}
\includegraphics[width=0.33\textwidth]{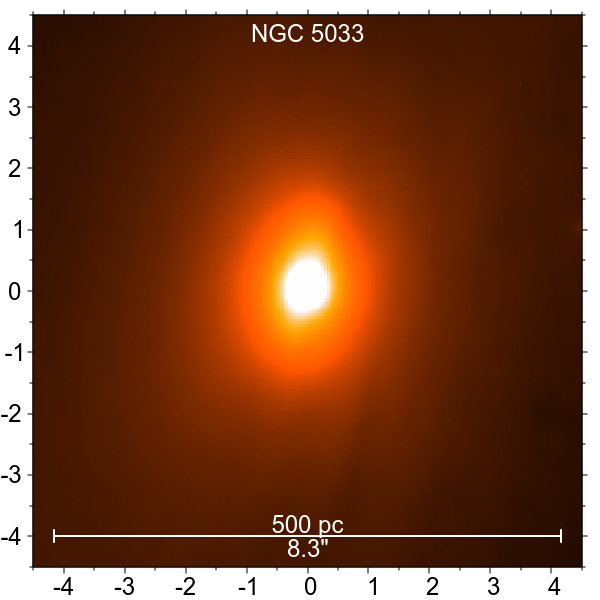}
\includegraphics[width=0.33\textwidth]{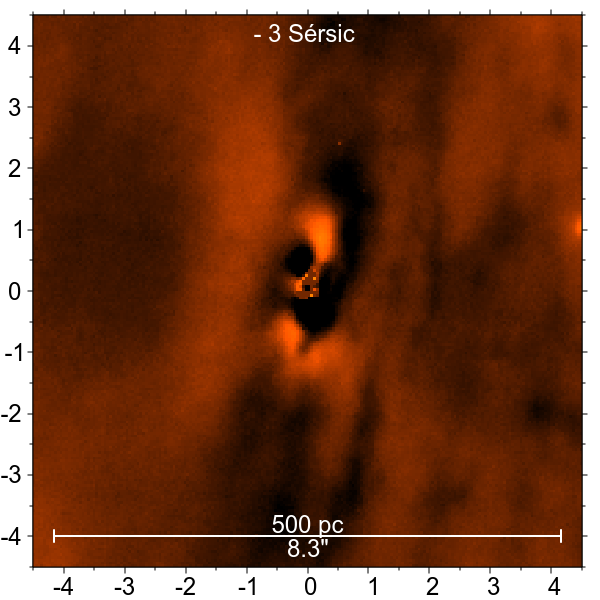}
\includegraphics[width=0.33\textwidth]{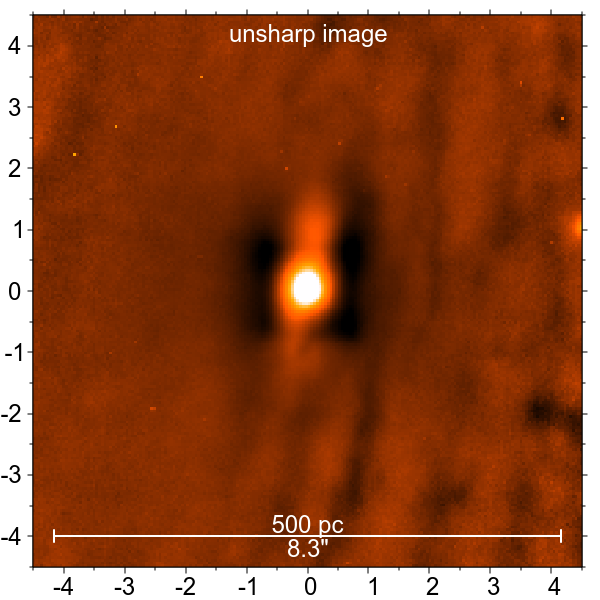}\\
\caption{Three representative objects of our sample. NGC~5695, which has a newly detected nuclear ring (top), NGC~3254, which has a (known) nuclear bar (center), and NGC~5033, which has no clear further galactic component (bottom). From left to right: Original image, best-fit image after the subtraction of 1--3 \ser profiles, and the unsharp-masked image.
North is up and east to the left on all images.}
\label{fitexample}
\end{figure*}

In four galaxies, we found clear evidence of an 
inner bar, which had already been reported in earlier studies in all four cases: NGC~3254 \citep{2015ApJS..219....4S}, NGC~3486 \citep{2013ApJ...771...59M}, NGC~4725 \citep{2004A&A...415..941E}, and NGC~4941 \citep{2000A&AS..145..425G,2002ApJ...567...97L,2004ApJ...616..707H,2024MNRAS.528.3613E}.
For three galaxies we detected a nuclear ring and all of them show at least one bar: NGC~4725 (6\farcs7 ring with a 
5\farcs6 inner and a 118" outer bar), NGC~4941 (2\farcs8 ring and a 3\farcs5 inner bar), and NGC~5695 (2\farcs1 ring and an 11"  large-scale bar). The bar measurements are from \cite{2004A&A...415..941E} for NGC~4725, and from \cite{2002ApJ...567...97L} for NGC~4941 and NGC~5695.
Two galaxies with a nuclear ring do not appear to host a bar at all, namely NGC~3607 and NGC~7466.
In Fig.~\ref{fitexample} we show the original image, the residual image after subtraction of the best-fit profile(s), and the unsharp-masked image for three representative objects: one showing a nuclear ring, one hosting a nuclear bar, and one without any obvious further 
galactic component.
The images for the full data set are presented in Appendix \ref{appendix1}. In Table~\ref{results} we summarize the properties known so far, present the results of our fits, compare them with the results obtained from the archival HST-band data, and highlight the main findings. 
Details of the fits to the HST data are given in Section~\ref{hstdata}.
In Appendix \ref{appendix2}, we comment on each object individually.
\begin{table*}[t]
\caption{Comparison of previously known and newly identified nuclear structures in the target galaxies.}
\centering
\vspace*{.2cm}
\begin{tabular}{l|c|cc|c}
\hline
\hline
\rule{0pt}{3ex}
Target   & Known properties                       & Our LBT results                                 & Our HST results   & What's new \\
\hline
\rule{0pt}{3ex}
MRK~461  & wound nuclear spiral-like              & tight spiral/ring 5\farcs75                           & tight wound spiral       & -\\
NGC~2985 & large-scale bar, 0\farcs5 ring         & nothing special                                      & nothing special           & - \\
NGC~3185 & large-scale ring, 2\farcs2 SF ring     & 2\farcs2 ring confirmed                               & ring                     & 2D model ring\\
NGC~3254 & narrow dust-lane, large-scale bar      & bar confirmed                                         & bar                      &  -\\
NGC~3486 & featureless, 25" bar                   & bar confirmed                                         & bar detected             & -\\
NGC~3607 & dust ring starts to develop            & 5\farcs8 ring-like structure                    & ring-like structure      & determination radius ring\\
NGC~4138 &  22" stellar ring and 4" gas ring      & no good fit                                           & disturbed center         & - \\
NGC~4725 & large-scale 11" and inner 5\farcs6 bar                    & bar confirmed, ring 6\farcs7                 & bar, ring        & detection ring\\
NGC~4941 &  3\farcs5 bar                          & bar confirmed, ring  2\farcs8                 & bar, ring        & detection ring\\
NGC~5033 &  triple bar? (uncertain)                   & nothing special                                       & whirly                   & - \\
NGC~5273 &  nuclear bar?                          & ambiguous                                              & ambiguous                & -\\
NGC~5347 &  nuclear dust spiral                   & 1" ring                                               & ring                     & detection / 2D model ring \\
NGC~5631 &  8" bulge                      & nothing special                                       & ring?                    & -\\
NGC~5695 &  dusty, large-scale bar                & 2\farcs1 ring                                       & ring                       &  detection ring \\
NGC~6211 &  nuclear shell, banana-shaped          & substructure, tidal tail?                             & banana-shaped            & -\\
NGC~7217 &  several rings                         & 2\farcs9 ring ?                                      & 4\farcs5 ring   ?           & new ring(s) ambiguous \\
NGC~7466 & not imaged at high-resolution before & 2\farcs6 ring                                    & no data                  & detection / 2D model ring\\
NGC~7674 & complex (bar, ring + spiral?)   & nothing special                                       & nothing special          & -\\
\hline
\end{tabular}
\tablefoot
{In column 2, we summarize the properties previously reported for the targets, based mostly on \textit{H} band and optical HST data. In the majority of cases, these properties were derived from the analysis of 1D surface-brightness profiles and/or
ellipse fits to the images to search for isophotal twists and related features. The relevant publications are listed in Appendix~\ref{appendix2}.
The properties of the targets derived from the LBT and HST data after subtraction of \ser profiles, visual inspection, and/or fitting of bars and ring-like structures to the residuals are given in Columns 3 and 4. With the exception of NGC~7217, which has outer, inner, and nuclear rings, "ring" always refers to a "nuclear ring". Where possible, we provide the radius or major axis of the ring. 
The previously unknown properties of the rings from our analysis are displayed in column 5.}
\label{results}
\end{table*}

\section{Discussion}

\subsection{Classification of detected rings and ring-like structures}
In the literature, a clear distinction is not always made between nuclear, inner, and outer rings.
\cite{1986ApJS...61..609B} used the term nuclear rings for very small structures that are often found in the centers of barred or weakly barred spirals. 
He used the term inner rings for luminous enhancements commonly observed around the bars of barred spirals or within the spheroidal component of nonbarred spirals. 
The outer rings, in turn, were defined as diffuse enhancements that envelope the main bodies of many lenticulars and early-type spirals.
Given the radii of the rings listed in Table~\ref{rings}, all of which are \(\leq 1.3\,\mathrm{kpc}\), the sample consists of nuclear or inner rings. 
The distinction between nuclear and inner rings is not always straightforward
\citep[e.g.][and references therein]{1996FCPh...17...95B,2010MNRAS.402.2462C}. However, it is generally known that inner rings tend to be found near the corotation radius of a bar; whereas nuclear rings are more often located at the inner end of a large-scale bar or the outer end of an inner bar. 
Distinguishing between a nuclear and an inner ring in the absence of a bar can be very difficult. 
Dynamically, nuclear rings are thought to lie just inside the ILR, which normally coincides with the inner end of the bar
\citep[e.g.][]{1984MNRAS.209...93S,1985A&A...150..327C}.
In unbarred galaxies, ILRs may instead be associated with spiral patterns, weak ovals, or perturbations induced by companion galaxies.

Six of the eight galaxies with rings in our sample host either a large-scale bar, a nuclear bar encircled by a ring, or both. 
These rings can therefore be unambiguously classified as nuclear rings. 
The two galaxies without evidence of a bar are NGC~3607 and NGC~7466. 
NGC~3607 forms a physical pair with NGC~3608 and is a member of a compact galaxy group \citep{2015MNRAS.446..120D}. The galaxy hosts a decoupled core of roughly the same size as the ring identified here. The former may be the result of either a minor merger or internal dynamical processes \citep{2007A&AT...26..311A}. 
Given these properties, and the relatively small radius of the ring in NGC~3607 (\(370\,\mathrm{pc}\)), we classified it as a nuclear ring. 
With a radius of \(1.3\,\mathrm{kpc}\), the ring in NGC~7466 is the largest in our small sample, although we note that this is not an uncommon size for nuclear rings \citep{2010MNRAS.402.2462C}. 
On deep legacy survey images\footnote{\url{https://www.legacysurvey.org/viewer}}, the galaxy appears asymmetric, perhaps even lopsided, and our image shows a lens-like central feature that may correspond to an inner bar (see Fig.~A.4). 
The diameter of the ring relative to the isophotal diameter of the galaxy, \(\log(D_{\mathrm{ring}} / D_{25}) = -1.54\), is typical of nuclear rings in Seyfert~2 galaxies (see Section \ref{ringprop}). 
Therefore, we also classify this ring as a nuclear ring.
Taken together, all rings in our sample are classified as nuclear rings, two of them in apparently unbarred galaxies.

\subsection{Regarding the high incidence of nuclear ring-like structures}

The unexpectedly large number of Seyfert~2 galaxies showing evidence of nuclear ring-like structures warrants closer examination.
At least 8 of the 18 galaxies in our sample (> 40\%) exhibit nuclear rings. 
We note that we did not detect a stellar nuclear ring in four galaxies that have been reported to host a nuclear ring in \cite{2010MNRAS.402.2462C}, namely NGC~2985, NGC~4138, NGC~5033, and NGC~7217. 
If these four galaxies are added to our eight detections, then 12 of the 18 galaxies in the sample (66\%) would be hosting either a star-forming or a stellar nuclear ring.
Since our observations were obtained in the \textit{K} band, and therefore primarily tracing an older stellar population, the nondetection of a star-forming ring would not be unexpected. 
Overall, this finding contrasts with earlier studies, which reported nuclear-ring features in at most 20\% of the sources \citep[e.g.][]{2010MNRAS.402.2462C}. 
In the following, we discuss possible reasons for this high incidence, including selection effects, wavelength coverage, and the advantages of carrying out a 2D analysis.

\subsubsection{Selection effects}

The sample selection is not expected to introduce a strong bias.
The targets were selected randomly based on observability during the observing runs, subject only to a redshift constraint imposed to ensure the highest possible spatial resolution. 
We did not preselect barred galaxies, which could otherwise have led to a higher detection fraction of nuclear ring-like structures, since bar-driven resonances can promote their formation. 
In fact, two of the eight galaxies with a ring in our sample show no evidence of a bar.
A chance coincidence cannot be ruled out entirely. However, under the assumption that 20\% of the 198 Seyfert~2 galaxies in the QSO catalog of \cite{2010A&A...518A..10V} satisfying our selection criteria host a nuclear ring, the probability of obtaining such a high detection rate by chance is only 1.1\%.

\subsubsection{Wavelength coverage: comparison with HST \textit{H}-band data}\label{hstdata}

As discussed in the introduction, \textit{Ks}-band observations are expected to be superior to \textit{H}-band observations for morphological studies of Seyfert~2 nuclei. 
Taking into account the effects of extinction, the larger aperture of the LBT compared to HST, the sky background seen by both telescopes, and the quantum efficiency of the detectors used, we estimate a gain in the signal-to-noise ratio (S/N) per pixel of at least a factor of 10 for our LBT \textit{Ks} band relative to the HST \textit{H}-band observations.

As can be seen from the fit results summarized in Table~\ref{results} and from an inspection of the residual images presented in Appendix \ref{appendix1}, all sources display essentially the same morphology regardless of the filters used.
This strongly supports the conclusion that the high incidence of nuclear rings or ring-like structures in our sample of Seyfert~2 galaxies is real. 
Contrary to expectations, our data show no clear advantage of the \textit{Ks} band over the \textit{H} band for this type of morphological analysis.

\subsubsection{2D analysis}
Since the sample is not expected to be strongly biased and since the gain from observing in the \textit{Ks} band (as compared to the \textit{H} band) appears to be limited, the differences likely arise from the analytical methodology. 
The 2D modeling of the images is superior to the analysis of radially extracted 1D surface brightness profiles. 
This is particularly important in cases where the components are faint and extended, as in our case, and/or when several galactic components, such as bars, are present.
The specific choice of 2D modeling with GALFIT or unsharp masking is less important, since the two approaches are complementary.
GALFIT has the advantage that it directly provides structural parameters such as sizes and fluxes, but it can fail to converge and is sensitive to the adopted number of components.
Unsharp masking is generally faster and more robust, but the choice of smoothing kernel becomes nontrivial when structures of different characteristic widths, such as bars and rings, are present.
To optimize the scientific return, it is best to use both methods in a complementary way, since this can help us to identify possible false detections. 
We encountered at least one instance where the 2D fits indicated a ring-like structure that was not supported by the unsharp-masked image (MRK~461) and another case showing the opposite behavior (NGC~5273).

Overall, PSF variations or similar effects are not expected to affect the fits significantly. 
The newly detected rings or ring-like structures, as well as those confirmed in this work, all have diameters \(> 1\arcsec\), corresponding to about 15 times the diffraction limit of the LBT in the \textit{K} band and at least six times the diffraction limit of the HST data.

\begin{figure*}[h]
 \centering

\includegraphics[width=0.25\textwidth]{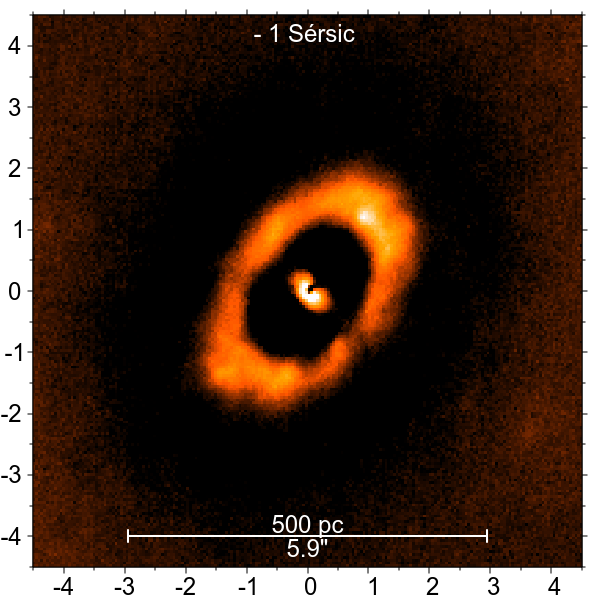}
\includegraphics[width=0.25\textwidth]{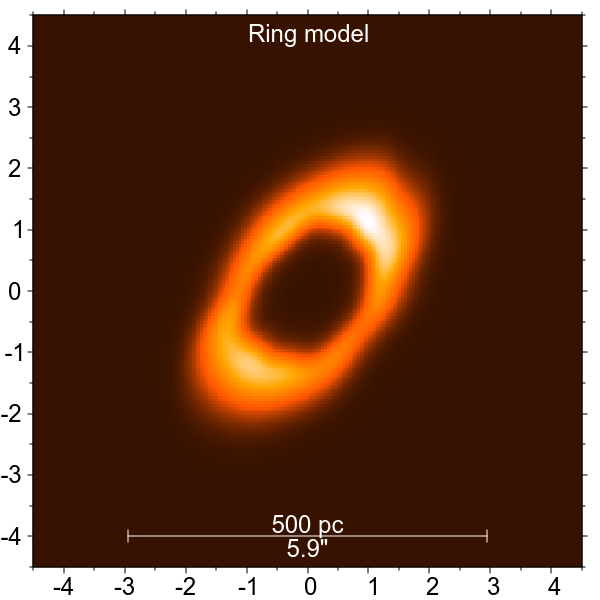}
\includegraphics[width=0.25\textwidth]{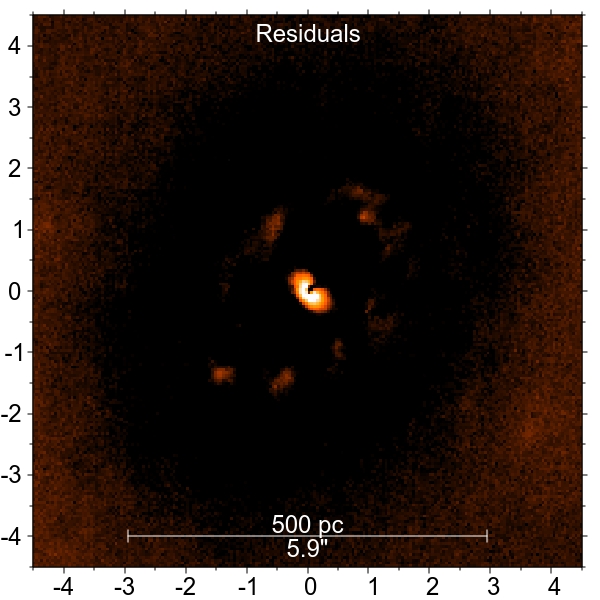}\\
\vspace*{.2cm}
\includegraphics[width=0.25\textwidth]{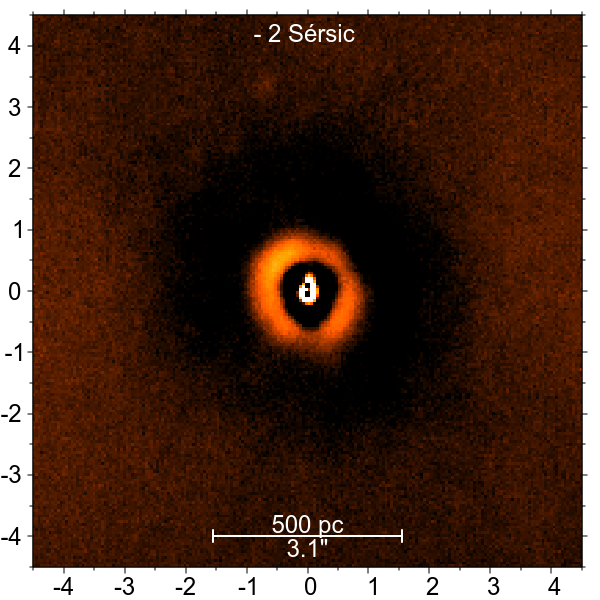}
\includegraphics[width=0.25\textwidth]{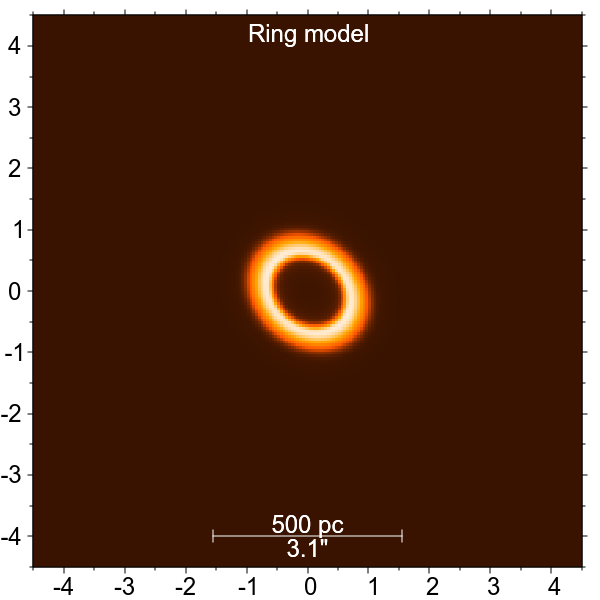}
\includegraphics[width=0.25\textwidth]{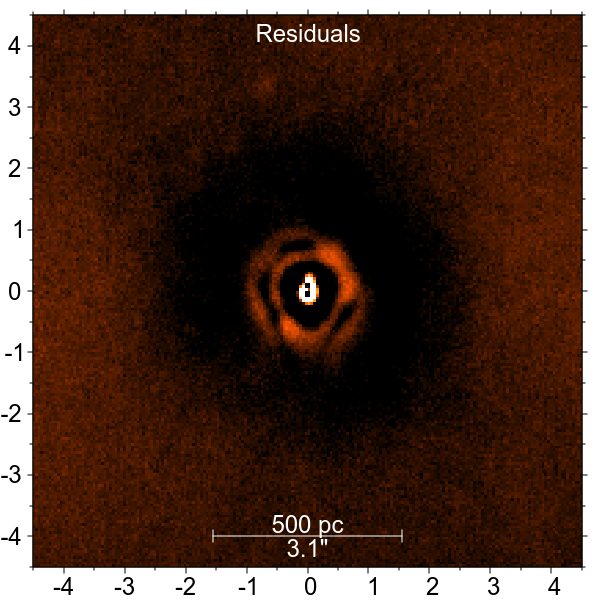}\\
\vspace*{.2cm}
\includegraphics[width=0.25\textwidth]{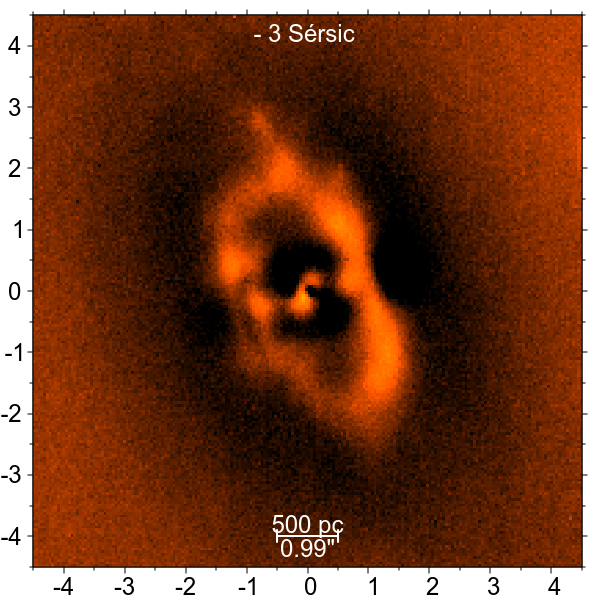}
\includegraphics[width=0.25\textwidth]{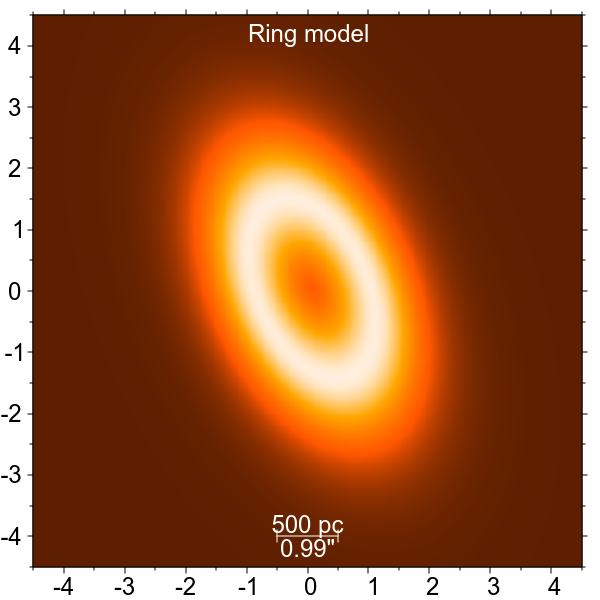}
\includegraphics[width=0.25\textwidth]{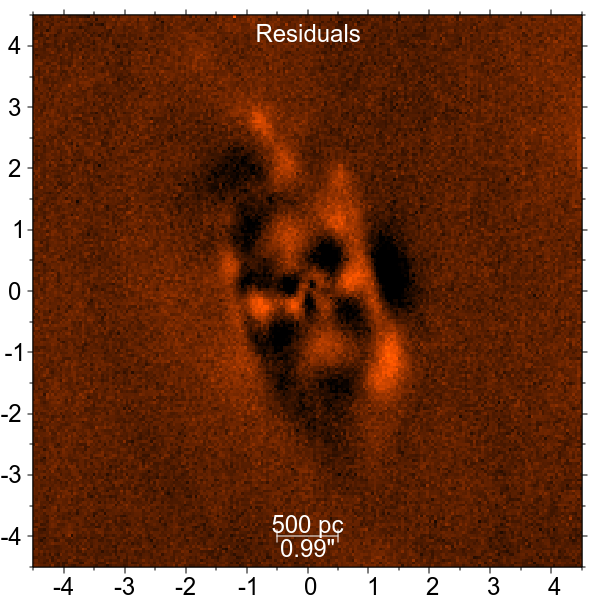}\\
\caption{Fits to the ring-like structures in the galaxies NGC~3185, NGC~5347, and NGC~7466.
From the left to right: Images after subtraction of \ser profiles, the best-fit ring models
and the residuals after subtraction of the ring component.
The FoV is \(9\arcsec \times 9\arcsec\). In all images, north is up and east is to the left.
}
\label{ringfit}
\end{figure*}
\subsection{Ring properties}\label{ringprop}
The ring properties were derived in two ways. 
First, we estimated their radii visually, which was possible for all sources.
Since not all sources displayed circular ring-like structures, the radii were measured along the major axis in the images. 
For three sources (NGC~3185, NGC~5347, and NGC~7466), we were also able to fit the 2D light distribution of the ring using GALFIT; the corresponding results are presented in Fig.~\ref{ringfit}. 
The radii and magnitudes from our fits to the three sources given above as well as the radii from visual estimates 
for the remaining five sources are listed in Table~\ref{rings}.

With the exception of NGC~7466, all ring-like structures in our sample have radii of \(< 0.7\,\mathrm{kpc}\) and three of them have radii \(\leq 0.21\,\mathrm{kpc}\).

Although Seyfert~2 galaxies have been observed in large numbers at high spatial resolution, dedicated searches for (or analyses of) nuclear ring-like structures in these systems remain relatively rare.  
Measurements of radii of ring-like structures can be found in \cite{2003ApJS..146..353M}, \cite{2010MNRAS.402.2462C} and \cite{2016IJAA....6..219A}. In addition, \cite{2003ApJS..146..353M} and \cite{2010MNRAS.402.2462C} presented an analysis of their own observations or archival (mostly HST) data, whereas \cite{2016IJAA....6..219A} compiled results on nuclear ring-like structures from the literature.

\begin{figure}[h]
 \centering
\includegraphics[width=0.49\textwidth]{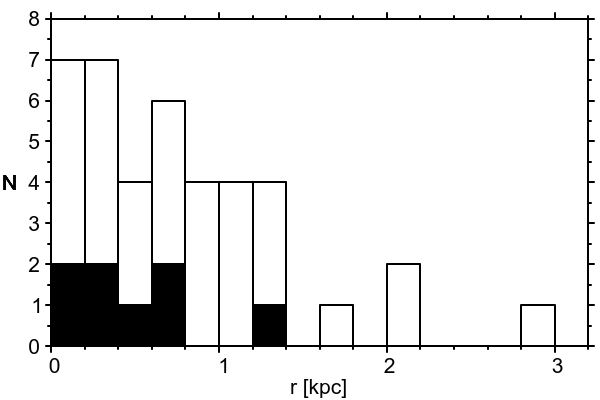}\\
\includegraphics[width=0.49\textwidth]{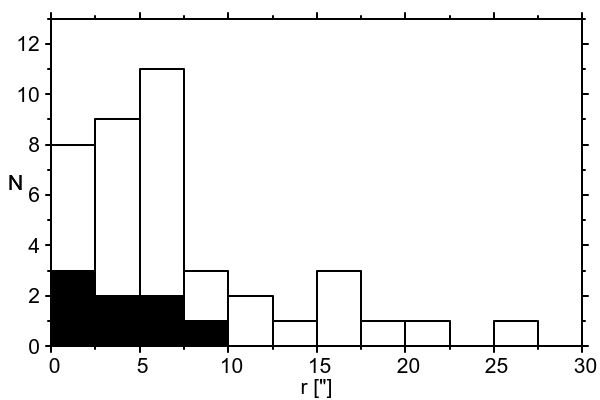}
\caption{Distribution of the ring radii of Seyfert~2 galaxies in kpc (top) and arcsec (bottom), 
comparing literature values (white area; see
\citealt{2003ApJS..146..353M,2010MNRAS.402.2462C,2016IJAA....6..219A}
and references therein) with those determined in this work
(black area).
The literature radii have been converted to the cosmology adopted here.}
\label{ringdim}
\end{figure}

Altogether, these studies derived ring properties for 32 Seyfert~2 galaxies. 
In Fig.~\ref{ringdim}, we compare the radius distribution of their Seyfert~2 galaxies with that of our sample.
As expected, our radii are clustered toward small values, which is not surprising given that our observations were obtained with an AO system and, therefore, they cover a comparatively small FOV.
Given the small number of sources, we did not attempt to separate the rings into systems dominated by dust, star-formation, or an old stellar population. 

We could ask whether the nuclear rings in Seyfert~2 galaxies differ in their properties from those in inactive galaxies, since this could point to a different formation scenario and/or to feedback from the active nucleus. 
This could be tested by comparing their normalized size distributions, \(D_{\mathrm{ring}} / D_{25}\), where \(D_{\mathrm{ring}}\) is the diameter of the ring and \(D_{25}\) is the isophotal diameter of the galaxy at a surface brightness level of \(25\,\mathrm{mag\,arcsec^{-2}}\).

\begin{figure}[h]
 \centering
\includegraphics[width=0.49\textwidth]{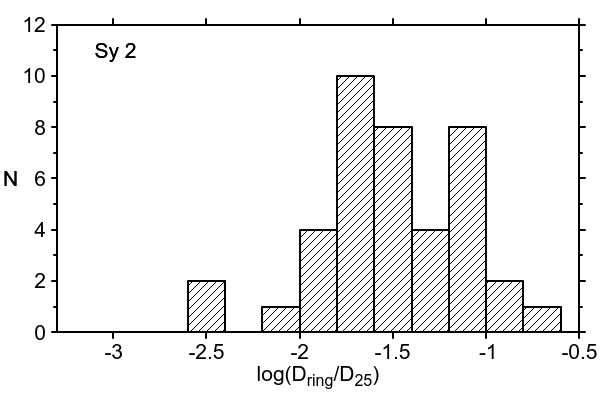}\\
\includegraphics[width=0.49\textwidth]{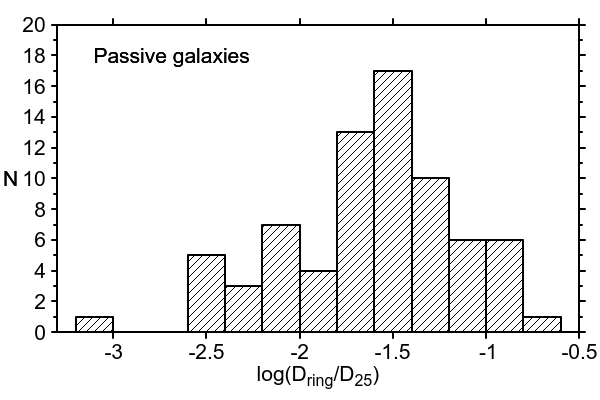}
\caption{Normalized size distribution of nuclear rings for 40 Seyfert~2 galaxies (top) and 73 passive galaxies (bottom).}
\label{ringcompare}
\end{figure}

By combining the properties of the 32 Seyfert~2 galaxies reported by \cite{2003ApJS..146..353M}, \cite{2010MNRAS.402.2462C}, and \cite{2016IJAA....6..219A} with our 8 ringed Seyfert~2 galaxies, thereby forming a sample of 40 objects, we derive \(\overline{\log(D_{\mathrm{ring}} / D_{25})} = -1.50 \pm 0.4\).
For the sample of 73 non-AGN galaxies with nuclear rings from \cite{2010MNRAS.402.2462C}, we derived \(\overline{\log(D_{\mathrm{ring}} / D_{25})} = -1.61 \pm 0.47\) with a median of \(-1.55\). The isophotal diameter, \(D_{25}\), is taken from the Hyperleda database for both the Seyfert~2 and the nonactive galaxy samples.
The two distributions are similar and are shown in Fig.~\ref{ringcompare}.
A Kolmogorov-Smirnov (KS) test does not indicate a significant difference between them ($p = 0.491$).

With the exception of NGC~3607, all nuclear rings in our sample are hosted by galaxies with RC3 \citep{1991rc3..book.....D} Hubble type \(T \sim 2\) (i.e., in the Sa--Sb range).
This is very similar to what has been found for the Seyfert~2 galaxies in the studies discussed above, where the majority of nuclear rings occur in galaxies with \(T = 3\) (Sb).
The same trend is seen in the sample of inactive (non-AGN) galaxies from \cite{2010MNRAS.402.2462C}, whose distribution also peaks at \(T = 3\). 
Thus, while nuclear rings may arise from inflowing material that can, in principle, feed an AGN, their formation does not appear to be directly coupled to AGN activity. 
Given that the properties of nuclear rings are largely indistinguishable between active and inactive galaxies, the occurrence of AGN activity appears to be independent of the existence of a nuclear ring.

\begin{figure}[h!]
 \centering
\includegraphics[width=0.49\textwidth]{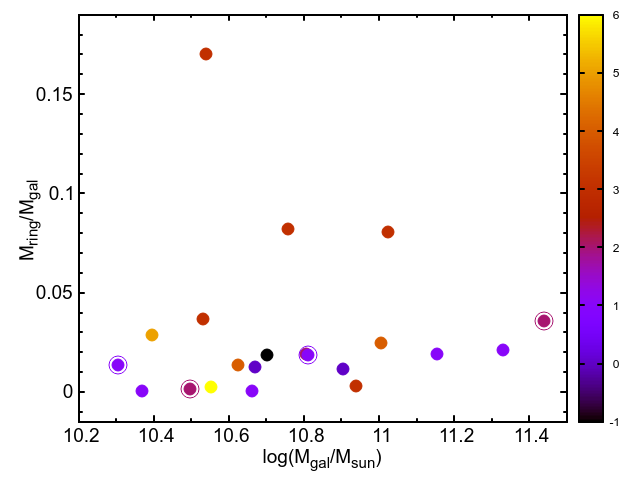}
\caption{Ratio of ring mass to total host-galaxy stellar mass, \(M_{\mathrm{ring}} / M_{\mathrm{gal}}\), as a function of host-galaxy stellar mass for our Seyfert~2 galaxies (NGC~3185, NGC~5347, and NGC~7466 as well as NGC~2273 from \citealt{2023AN....34430094S}, encircled) and the non-active galaxies from \cite{2018ApJ...857..116M}, color-coded by Hubble type.}
\label{ringmass}
\end{figure}

\begin{table}
\caption{Properties of the detected nuclear rings and ring-like structures.}
   \centering
    \begin{tabular}{l|rrc}
    \hline
    \hline
    \rule{0pt}{3ex}
Target &  $r_A$  & $r_A$     & $m_{K\mathit{s}}$ \\ 
       & ["] & [kpc] & [mag] \\
       \hline
       \rule{0pt}{3ex}
NGC~3185 & 2.15 & 0.18 & 13.86\\
NGC~3607 & 5.80 & 0.37 & \\       
NGC~4725 & 6.70 & 0.56 & \\
NGC~4941 & 2.80 & 0.21 & \\ 
NGC~5347 & 1.00 & 0.16 & 17.00 \\
NGC~5695 & 2.10 & 0.61 & \\ 
NGC~7217 & 10.0 & 0.66 & \\
NGC~7466 & 2.60 & 1.31 & 13.53\\
    \end{tabular}
    \tablefoot{Column 1 gives the target, and columns 2 and 3 give the radius of the ring or ring-like structure along its major axis, 
    in arcsec and kpc, respectively. Column 4 lists the brightness of the ring derived from our fits.
    }
    \label{rings}
\end{table}

\subsection{Stellar masses of the rings}\label{ringmass1}

We estimated the ring masses for the galaxies in which the rings could be modeled, following the prescriptions of \cite{2023AN....34430094S}. 
This resulted in masses of \(2.8 \times 10^8\), \(4.0 \times 10^7\), and \(9.8 \times 10^9\,M_\odot\) for the nuclear rings in NGC~3185, NGC~5347, and NGC~7466, respectively.
After including NGC~2273 (\(1.2 \times 10^9\,M_\odot\); \citealt{2023AN....34430094S}) among these galaxies, we were able to estimate the masses of the nuclear rings in four Seyfert~2 galaxies, which lie in the same range as the masses found for nuclear rings in 17 inactive galaxies by \cite{2018ApJ...857..116M}. 
Their mass range extends between \(1.3 \times 10^7\) and \(8.5 \times 10^9\,M_\odot\).
We caution that the methods used to derive the stellar mass of the rings are different and are subject to uncertainties, as discussed by \cite{2018ApJ...857..116M} and \cite{2023AN....34430094S}.
To compare the stellar masses of the rings to the stellar masses of their hosting galaxies we estimated galaxy stellar masses 
for both the \cite{2018ApJ...857..116M} and our galaxies, following the discussion in \cite{2003ApJS..149..289B}. This was achieved by using the color-dependent \textit{K}-band mass-to-light ratios, \((M_{\mathrm{gal}}/L) = \alpha_{\mathit{K}} + \beta_{\mathit{K}} \times (B - V)\), given in solar units. The \textit{K}-band luminosities, the corresponding $B-V$ colors, and the luminosity distances (modbest) were obtained from the HyperLeda database. The \textit{K}-band extinctions were taken from the NED. 
For the absolute solar luminosity in the \textit{K} band, we adopted \(M_{\mathit{K}} = 3.27\) \citep{2018ApJS..236...47W}.

In Fig.~\ref{ringmass}, we show the ratio of ring mass to total stellar mass of the host galaxy, \(M_{\mathrm{ring}} / M_{\mathrm{gal}}\), as a function of \(\log(M_{\mathrm{gal}})\), color-coded based on the revised Hubble type, for our galaxies and for those from \cite{2018ApJ...857..116M}.
As can be seen, a substantial fraction of the stellar mass of these galaxies appears to reside in their nuclear rings.
We identified the same trend reported in \cite{2018ApJ...857..116M}, namely, an increasing ratio of ring mass to total stellar mass with increasing host-galaxy stellar mass, with our objects extending the relation toward both the lowest and highest host-galaxy masses. 
Again, the nuclear rings of Seyfert~2 galaxies do not differ in any obvious way from those in non-active galaxies. 
We derived somewhat higher host-galaxy masses than \cite{2018ApJ...857..116M} obtained for their sample, but suspect that this is due to the different sources used for extracting the luminosity distances (NED versus Hyperleda). 
We also see evidence that the stellar ring mass fraction relative to the host-galaxy stellar mass is highest for \(T = 3\) (Sb-type) galaxies, although this trend is less pronounced than in \cite{2018ApJ...857..116M}.

\subsection{Considering whether ring size is correlated with the galaxy potential}

As discussed in the previous sections, nuclear rings can be formed via bars and related Lindblad resonances. 
Although details are subject to several intrinsic properties of the galaxy (e.g., bar pattern speed), we would naively expect the radius of a nuclear ring to scale with the mass of the galaxy, since the location of the inner Lindblad resonance and the extent of the \(x_{2}\) orbital family are governed by the underlying gravitational potential and rotation curve (e.g., \citealt{2018MNRAS.481....2S,2024MNRAS.528.5742S}). 
To explore this possibility, we examined the 40 Seyfert~2 galaxies with measured ring radii presented in Section~\ref{ringprop}, using their host-galaxy stellar masses (derived in Section~\ref{ringmass1}) as a proxy for the potential depth. 
When considering the entire sample, we found only a weak and statistically marginal correlation, with a linear slope of 0.35 for the relationship between the radius of the ring and \(\log M_\star\), along with Pearson and Spearman coefficients of \(r = 0.291\) (\(p = 0.0685\)) and \(\rho = 0.242\) (\(p = 0.133\)), respectively.
A similarly weak trend was reported by \cite{2024MNRAS.528.3613E} for barred galaxies.
However, when restricting the analysis to the eight Seyfert~2 galaxies for which we have homogeneous ring measurements, a much clearer correlation emerges. 
For this subsample, we obtained a slope of 0.85 with $r = 0.740$ ($p = 0.00929$) and $\rho = 0.700$ ($p = 0.0164$). 
Given the small number of objects and the fact that the subsample still includes a mix of barred and nonbarred galaxies, the correlation should be interpreted with caution. 
This might reflect the particular characteristics of this restricted set rather than a universal relation for all nuclear rings.

\subsection{Ring formation and timescales}

Our analysis in previous sections has shown that the properties of nuclear rings in low-luminosity active galaxies do not differ from those of their counterparts in non-active galaxies. 
Apparently, low-level AGN activity does not, at least, prevent the formation of nuclear rings. 
The question remains, however, why we find that 40\% or more of our sources host a nuclear ring, a fraction considerably higher than the 10\% derived by \cite{1998ApJS..117...25M} for Seyfert~2 galaxies and the 20\% estimated by \cite{2010MNRAS.402.2462C} for disk galaxies based on much larger samples.

We start by assuming that our results reflect a genuine physical effect; namely, that the larger fraction of Seyfert~2 galaxies with nuclear rings is of a physical origin and that neither one is the result of a bias, nor a chance coincidence.
There are various channels that can support the formation of nuclear rings. 
In all cases, a perturbation of the triaxial potential in the central region of the galaxy is required. 
The most obvious and probably most important mechanism is described, for example, by \cite{1999ASPC..187..100S} and \cite{2001ASPC..249...55S}. 
In this scenario, the perturbation is induced by a galactic bar, which drives gas inward along shocks.
When the shocked gas crosses an ILR, it can become trapped, and the trajectories shift from radial to nearly circular, eventually forming a molecular ring. 
Once the molecular gas becomes sufficiently dense, star formation might begin and not necessarily uniformly across the ring.

This scenario is supported by a number of IFU observations of nuclear rings in passive 
and low-luminosity AGNs, which could be dominated by young (\(\leq 10\,\mathrm{Myr}\)) starbursts 
\citep[e.g.,][]{2017A&A...598A..55B,2019A&A...622A.128F,2020A&A...638A..53F,2020A&A...638A..36F}
and intermediate-age (\(100\text{--}700\,\mathrm{Myr}\)) stellar populations 
\citep[e.g.,][]{2010ApJ...713..469R,2011MNRAS.417.2752R,2017MNRAS.469.3286D,2017MNRAS.470..992R},
whose contribution to the continuum in the \textit{K} band can be up to 100\% (\citealt{2017MNRAS.469.3286D}). 
\cite{2018ApJ...857..116M} derived typical ages of \(100\,\mathrm{Myr}\) for the individual stellar clusters in nuclear rings, and an averaged age of \(\sim 1.2\,\mathrm{Gyr}\) from broadband SED fits to the integrated rings of 18 inactive disk galaxies. 
They argued that nuclear rings can be long-lived structures experiencing multiple episodes of strong star formation. 
The oldest stellar population so far identified in a nuclear ring, with a mean age of \(\sim 2.7\,\mathrm{Gyr}\), was found in the Seyfert~2 galaxy NGC~5806 by \cite{2025ApJ...980...64R}.
Similar lifetimes for nuclear rings were estimated by \cite{2010MNRAS.402.2462C} on statistical grounds.\\
The typical lifetimes of bars remain far from clear. 
\cite{2005MNRAS.364L..18B} argued, based on numerical simulations, that bars may be transient phenomena with typical lifetimes of \(1\text{--}2\,\mathrm{Gyr}\). 
More recently, \cite{2025A&A...698A...5D} derived bar formation epochs, or bar ages, for nearby disk galaxies in their TIMER sample ranging from \(1\) to \(13.5\,\mathrm{Gyr}\) ago depending on their star formation rate surface densities (SFD). High-SFD galaxies and low-SFD-galaxies have typical bar ages of about \(4\,\mathrm{Gyr}\) and about \(9\,\mathrm{Gyr}\), respectively. Taken together, the characteristic lifetimes of bars and rings could differ by 
a factor of 2 to 4.

Depending on whether the detection is made in the optical or near-infrared, and on whether weak bars are included in the statistics, the fraction of barred galaxies among spirals in the local Universe has been found to lie between 26\% \citep{2016A&A...595A..63V} and 60\% \citep{2018MNRAS.474.5372E} for our covered galaxy stellar mass range.
Under the assumption that every barred galaxy develops a nuclear ring and that both appear only once during the lifetime of the galaxy, 
the expected fraction of barred Seyfert~2 spiral galaxies hosting nuclear rings could be somewhere between 7\% and 30\%.

Estimating the fraction of nonbarred Seyfert~2 galaxies hosting a nuclear ring is much more difficult. Rings in these
galaxies may form through ovals, strong (one-sided) spiral arms, lopsidedness, or interactions with other galaxies 
(e.g., \citealt{2001sac..conf..223C}), the relative number of each of these is tricky to determine. Attempts to determine numbers 
have been made by, for instance, 
\cite{1995ApJ...447...82R} and \cite{1997ApJ...477..118Z}, who found that at least 30\% of spiral galaxies show asymmetric stellar distributions, 
and by \cite{1994A&A...290L...9R}, who found that at least 50\% of nonbarred galaxies show asymmetric H\,{\sc i} profiles. 
The fraction of galaxies displaying evidence of tidal interaction has been determined by \cite{2025MNRAS.544..735S} to be at least 12\%. 
The timescales associated with ovals, strong (one-sided) spiral arms, lopsidedness, or interactions vary, but can be longest
for lopsidedness maintained by persistent gas accretion by the cosmic environment. \cite{2005A&A...438..507B} estimated that this process 
can last for several Gyrs for the latter. If we assume a timescale of \(5\,\mathrm{Gyr}\) for lopsidedness inducing a ring and again a timescale of \(2.7\,\mathrm{Gyr}\)
for the lifetime of a ring (i.e., a difference in their lifetimes by a factor of 2) and fold in that about 40\% of all spiral galaxies
are unbarred, we find that at most 20\% of nonbarred Seyfert~2 spiral galaxies could host a nuclear ring.

\section{Summary and conclusions}

In this paper, we present a 2D analysis using GALFIT, along with unsharp-masked images, of the nuclei of a brightness- and redshift-limited sample of 18 randomly selected Seyfert~2 galaxies. 
The AO-aided images used for this analysis were obtained in the \textit{Ks} band with the LBT.
In addition, we used \textit{V}- or \textit{H}-band images of the same sources taken with HST to compare the nuclear morphologies. 
These images were cleaned of the superimposed smooth 2D surface-brightness component representing the disk and bulge.

The motivation for this study was twofold: (a) to test whether \textit{Ks}-band images are superior to images obtained at shorter wavelengths for the study of the nuclei of active disk galaxies, where extinction can be significant, and (b) to assess whether a 2D analysis offers advantages over a 1D approach, which has been used in most, though not all, prior studies.

Our main findings are as follows:
\begin{itemize}
\item[-] Independent of the band used, the nuclear morphologies are basically identical for each source once the bulge and disk components have been removed. 
Thus, at least for our sample, there is no clear advantage in moving to the \textit{Ks} band.
\item[-] A 2D analysis is valuable because it reveals low-luminosity features that are not detectable~in 1D radial surface brightness profiles.
\item[-] The complementary use of GALFIT (or a similar fitting tool) and unsharp masking (or related techniques such as structure maps) is highly advantageous. 
Most importantly, it helps to identify false positives; namely artifacts produced either by poor fits in GALFIT or by an inappropriate choice of kernel in unsharp masking. 
The two methods are technically different: while unsharp masking depends primarily on the adopted kernel, GALFIT is highly sensitive to the input parameters in multicomponent fits. 
It is therefore unlikely that both methods would produce the same incorrect result. 
In addition, GALFIT enables the derivation of physical parameters of the individual components beyond their sizes.
\item[-] Notably, we found a nuclear ring in at least 8 of the 18 galaxies in our sample (\(44 \pm 12\%\); for the determination of the 
error we assumed a binomial distribution).
This is about a factor of two higher than reported in the most detailed study by \cite{2010MNRAS.402.2462C}.
Five of these are new detections. 
For four of them, HST data exist, but the nuclear rings were not recognized in earlier studies.  
\item[-] We compared the properties of the rings (radii, stellar masses) and the morphological types of their hosting galaxies with those in inactive galaxies and found no significant differences. 
Low-luminosity nuclei therefore appear to have little or no effect on the formation and evolution of nuclear rings.
\end{itemize}
A natural next step would be to analyze a large sample of high-resolution NIR images using a 2D decomposition complemented by unsharp masking, as done in the present study. 
Only this type of approach would enable the reliable detection of very faint nuclear rings.
While \textit{K}-band observations are ideal for minimizing extinction, the results given in Section~\ref{hstdata} demonstrate that \textit{H}-band data are sufficient for this analysis.
A reanalysis of HST \textit{H}-band snapshot observations of 250 galaxies, containing carefully matched samples of Seyfert 1, Seyfert~2, LINER, starburst, and normal galaxies presented by \cite{2004ApJ...616..707H}, could therefore help to confirm or refute our findings.

\begin{acknowledgements}
We thank the referee for their critical and detailed comments, which significantly improved the presentation of this paper.
The LBT is an international collaboration among institutions in the United States, Italy and Germany. LBT Corporation partners are: The University of Arizona on behalf of the Arizona Board of Regents; Istituto Nazionale di Astrofisica, Italy; LBT Beteiligungsgesellschaft, Germany, representing the Max-Planck Society, The Leibniz Institute for Astrophysics Potsdam, and Heidelberg University; The Ohio State University, and The Research Corporation, on behalf of The University of Notre Dame, University of Minnesota and University of Virginia.
This work was supported in part by the German federal department for education and research (BMBF) 
under the project numbers 05 AL2VO1/8, 05 AL2EIB/4, 05 AL2EEA/1, 05 AL2PCA/5, 05 AL5VH1/5, 05 AL5PC1/1 and 05 A08VH1. 
FPN gratefully acknowledges the generous and invaluable support of the Klaus Tschira Foundation.
FPN acknowledges funding from the European Research Council (ERC) under the European Union's Horizon 2020 research and innovation program (grant agreement No 951549).
This research is based on observations made with the NASA/ESA {\it Hubble} Space Telescope obtained from the Space Telescope Science Institute, which is operated by the Association of Universities for Research in Astronomy, Inc., under NASA contract NAS 5–26555. These observations are associated with programs 5446, 7328, 7330, 7856, 8597, 9360, 11080, 11219 and 13324.
We acknowledge the usage of the HyperLeda database (http://leda.univ-lyon1.fr). 
We finally  would like to thank the AO and TO-groups at the LBT for their excellent support during the observations.
\end{acknowledgements}

\bibliographystyle{aa}

\bibliography{Sy2AO}
\newpage
\begin{appendix}
\onecolumn

\section{Fits to LBT and HST data.}\label{appendix1}

\begin{figure}[h]
 \centering
\includegraphics[width=0.24\linewidth]{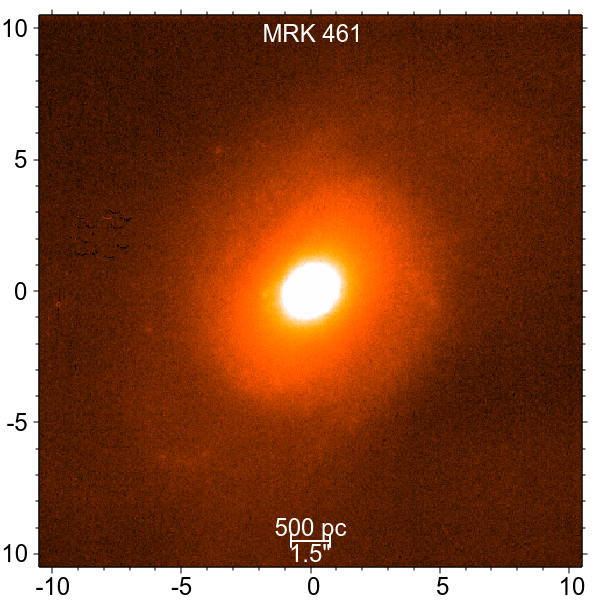}
\includegraphics[width=0.24\linewidth]{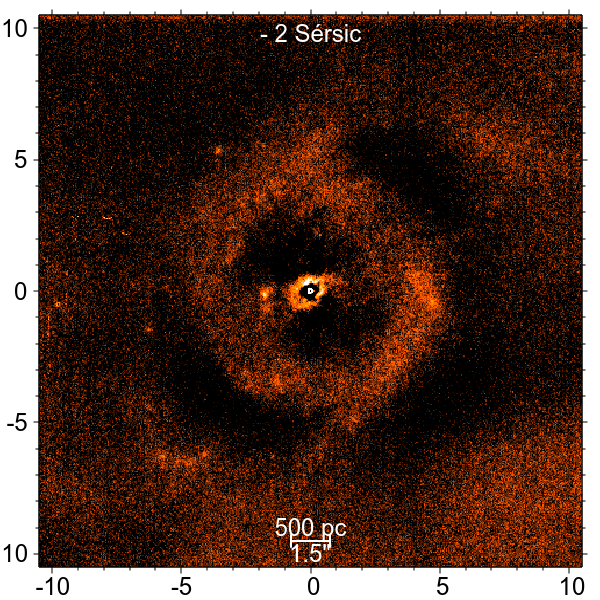}
\includegraphics[width=0.24\linewidth]{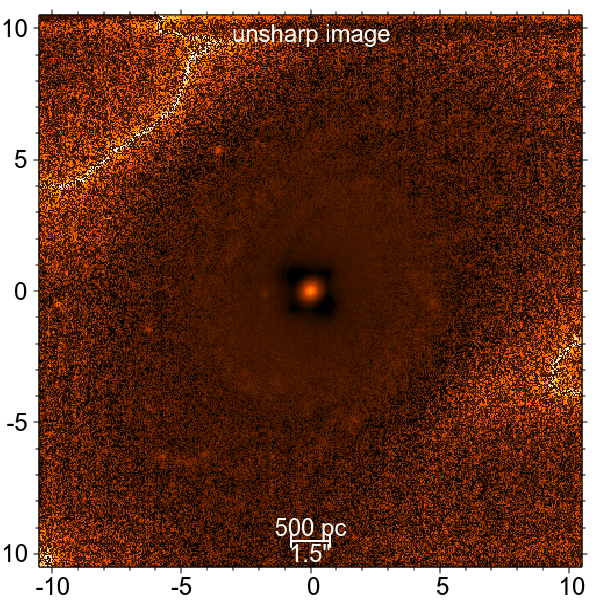}
\includegraphics[width=0.24\linewidth]{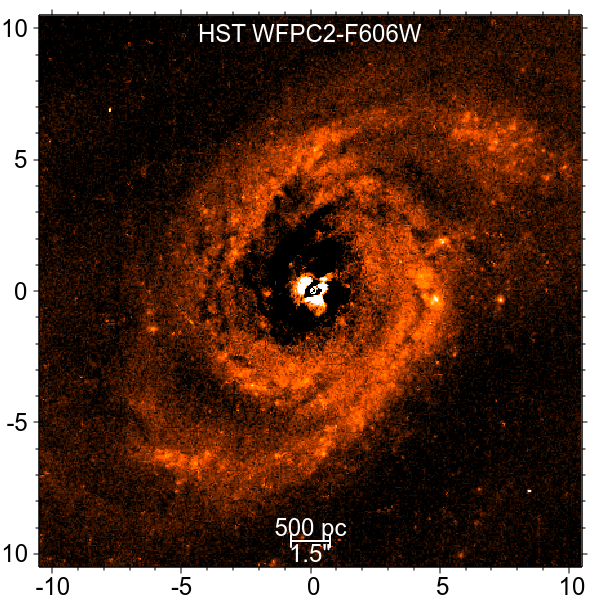}\\
\vspace*{.2cm}
\includegraphics[width=0.24\linewidth]{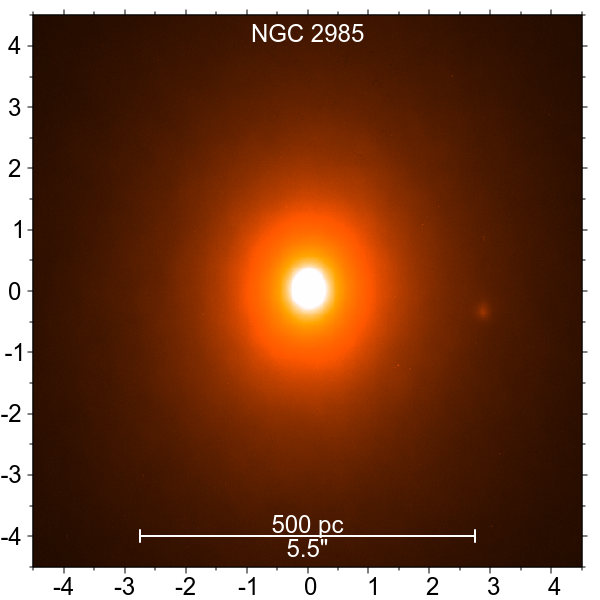}
\includegraphics[width=0.24\linewidth]{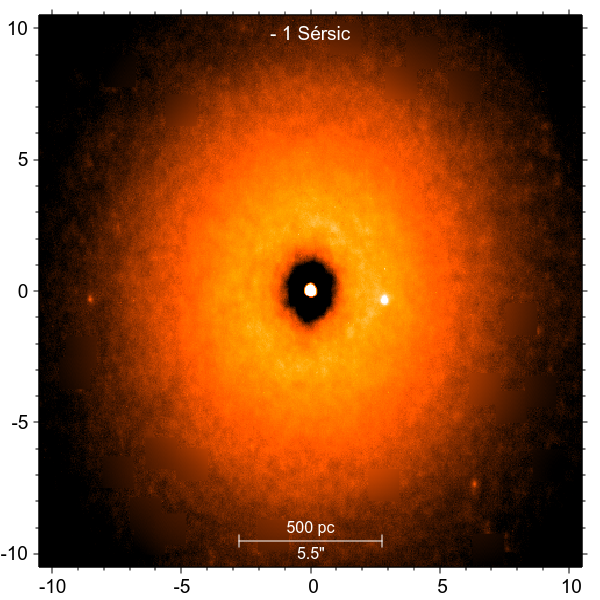}
\includegraphics[width=0.24\linewidth]{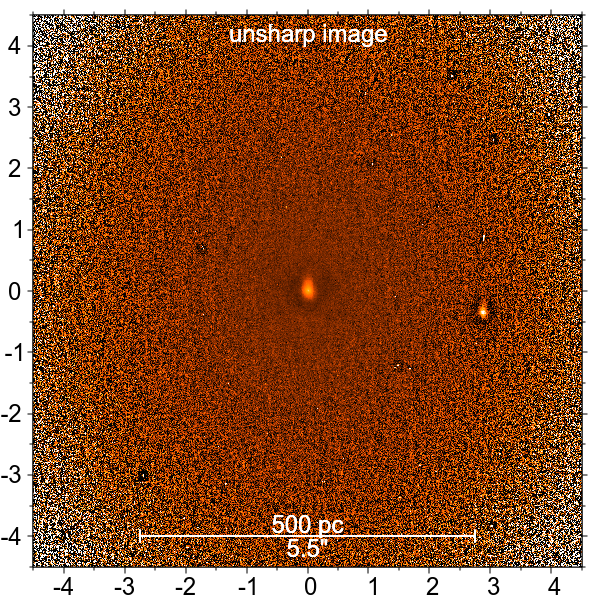}
\includegraphics[width=0.24\linewidth]{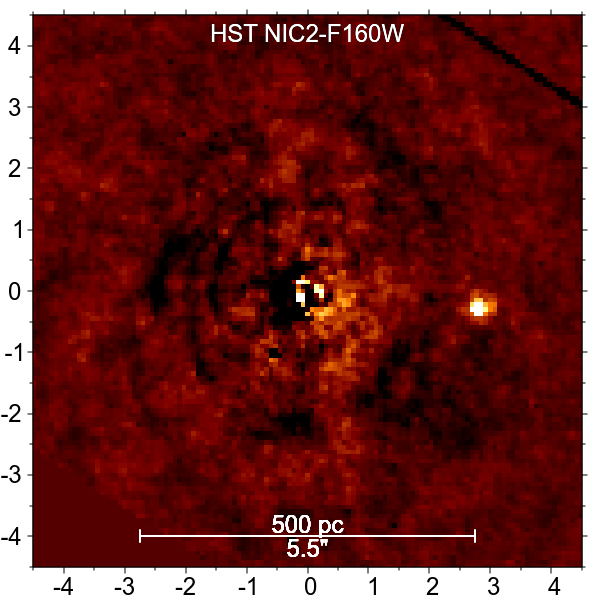}\\
\vspace*{.2cm}
\includegraphics[width=0.24\linewidth]{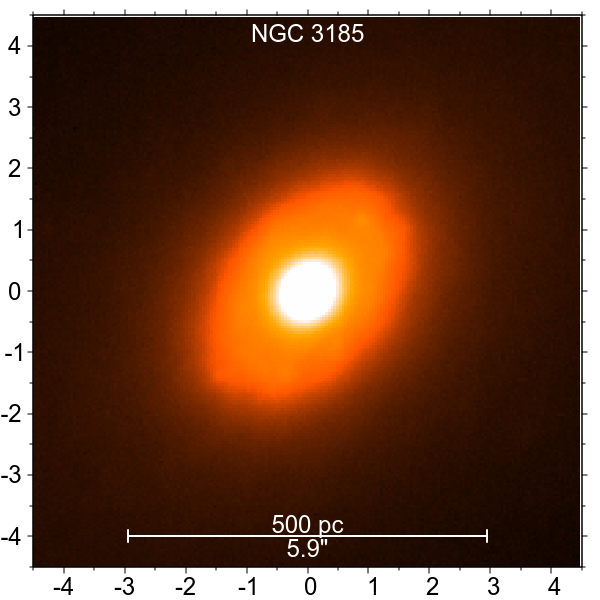}
\includegraphics[width=0.24\linewidth]{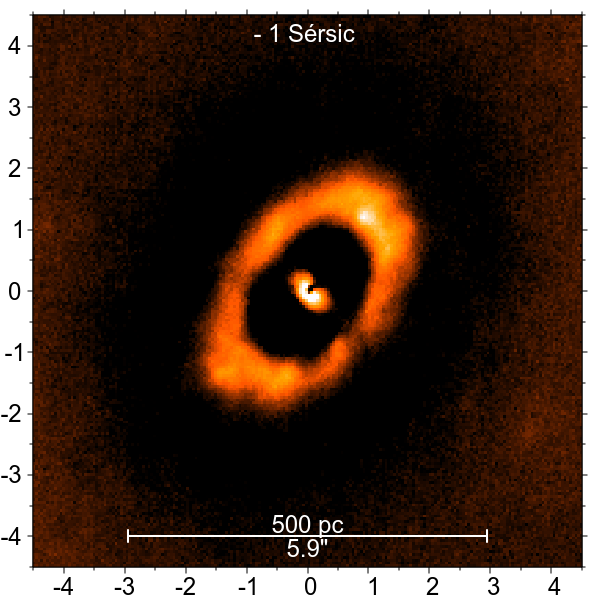}
\includegraphics[width=0.24\linewidth]{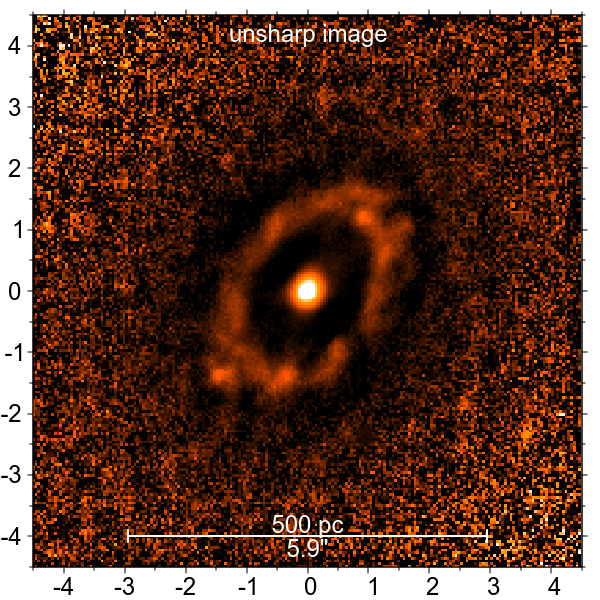}
\includegraphics[width=0.24\linewidth]{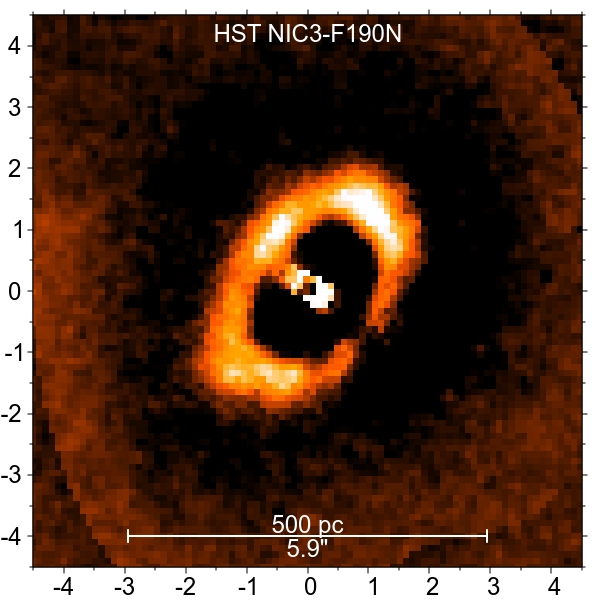}\\
\vspace*{.2cm}
\includegraphics[width=0.24\linewidth]{pics2/raltorig3254.png}
\includegraphics[width=0.24\linewidth]{pics2/raltfit3254.png}
\includegraphics[width=0.24\linewidth]{pics2/raltunsh3254.png}
\includegraphics[width=0.24\linewidth]{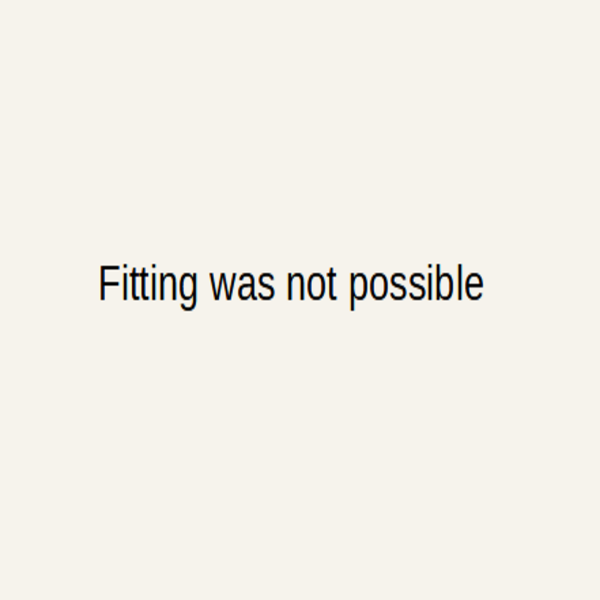}\\
\caption{Analysis of the images of  MRK~461, NGC~2985, NGC~3185, and NGC~3254. From left to right, we show the original \textit{Ks}-band image, the residual image after subtraction of one to three \ser components, the unsharp-masked image, and the residual image obtained after subtraction of one to three \ser components from the HST image. The FoV, scale, and the filter used for the HST images are indicated. North is up and east to the left in all images.}
\label{allfits}
\end{figure}

\begin{figure*}
 \centering
\includegraphics[width=0.24\linewidth]{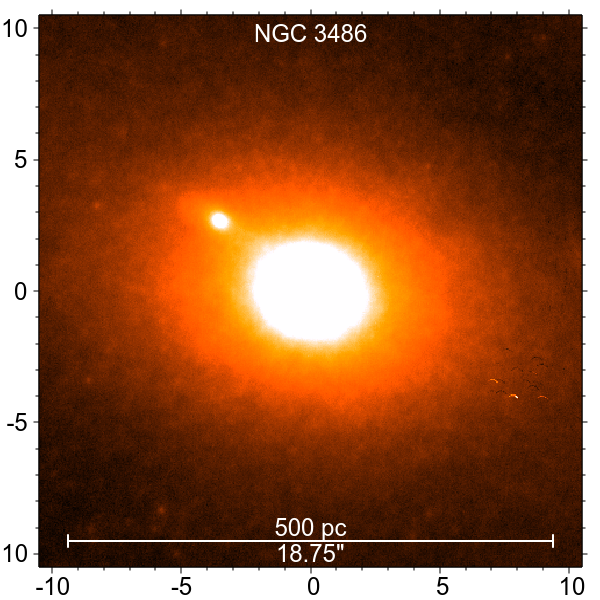}
\includegraphics[width=0.24\linewidth]{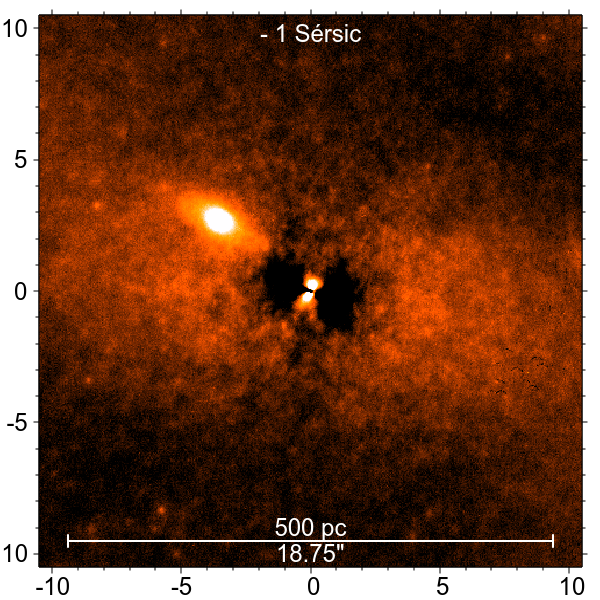}
\includegraphics[width=0.24\linewidth]{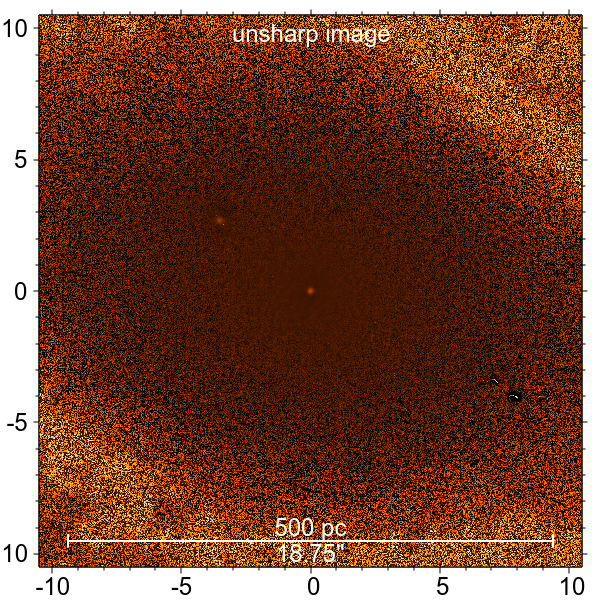}
\includegraphics[width=0.24\linewidth]{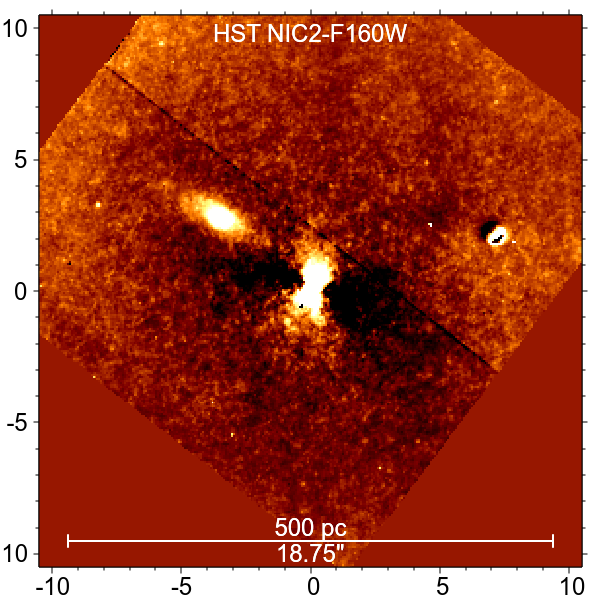}\\
\vspace*{.2cm}
\includegraphics[width=0.24\textwidth]{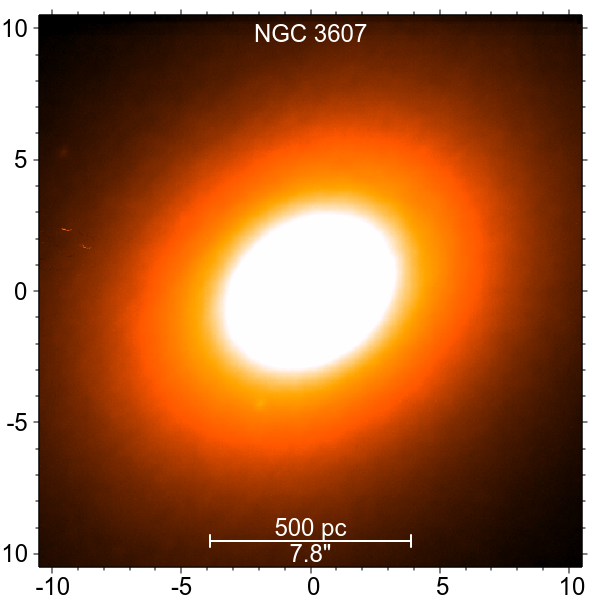}
\includegraphics[width=0.24\textwidth]{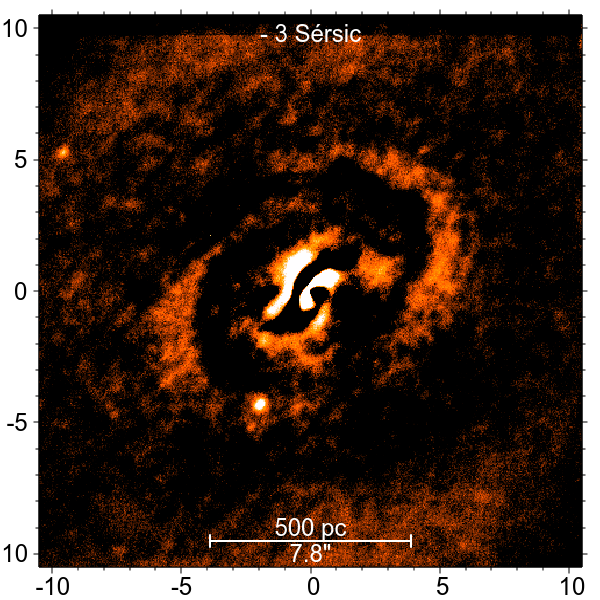}
\includegraphics[width=0.24\textwidth]{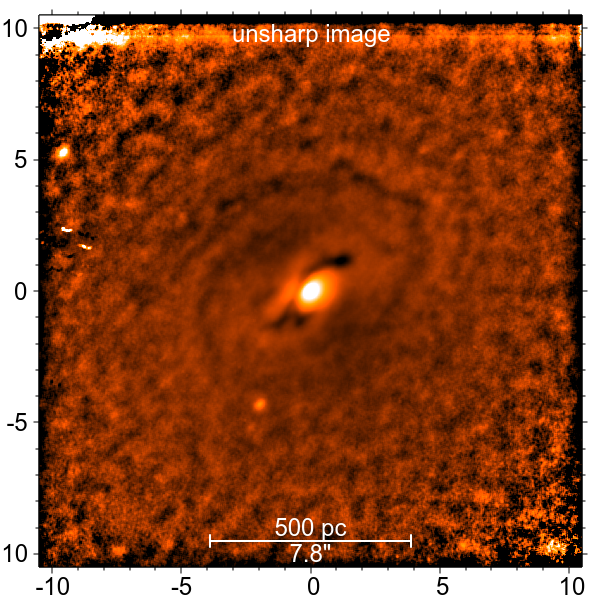}
\includegraphics[width=0.24\textwidth]{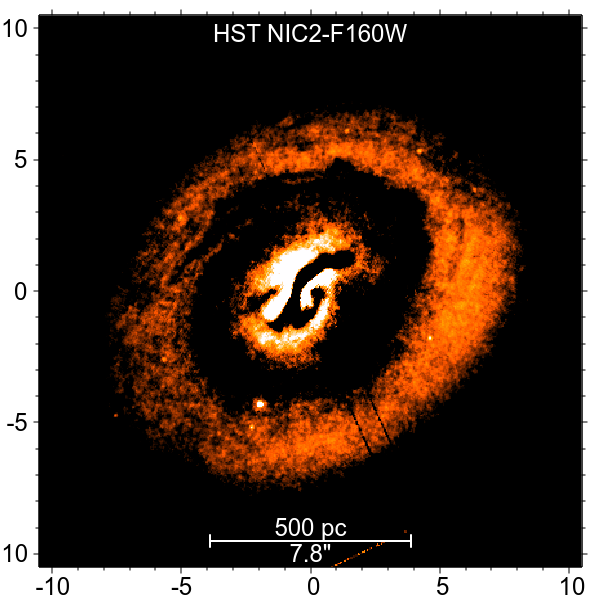}\\
\vspace*{.2cm}
\includegraphics[width=0.24\textwidth]{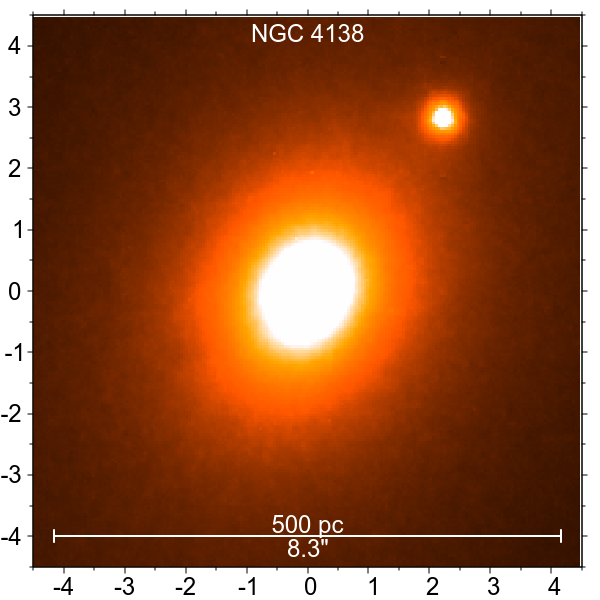}
\includegraphics[width=0.24\textwidth]{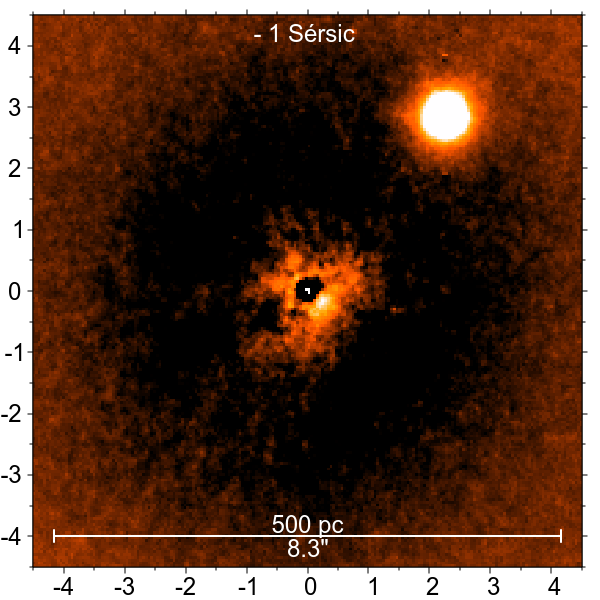}
\includegraphics[width=0.24\textwidth]{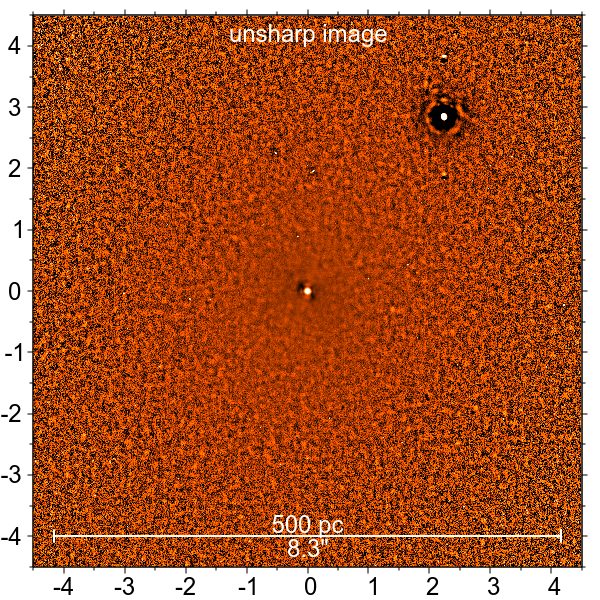}
\includegraphics[width=0.24\textwidth]{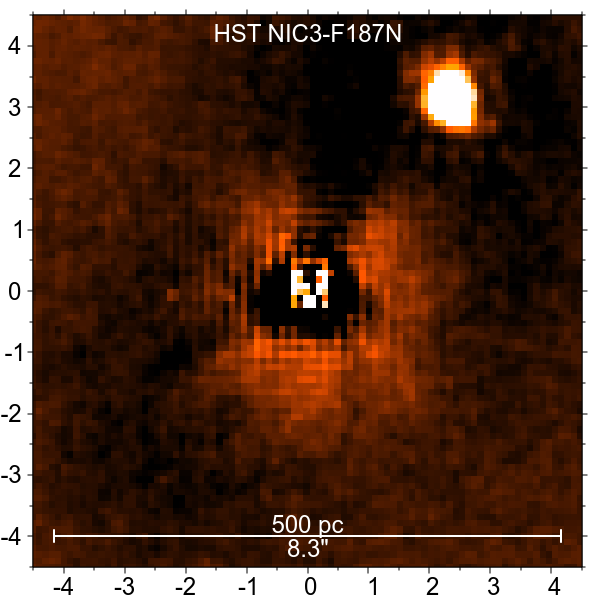}\\
\vspace*{.2cm}
\includegraphics[width=0.24\textwidth]{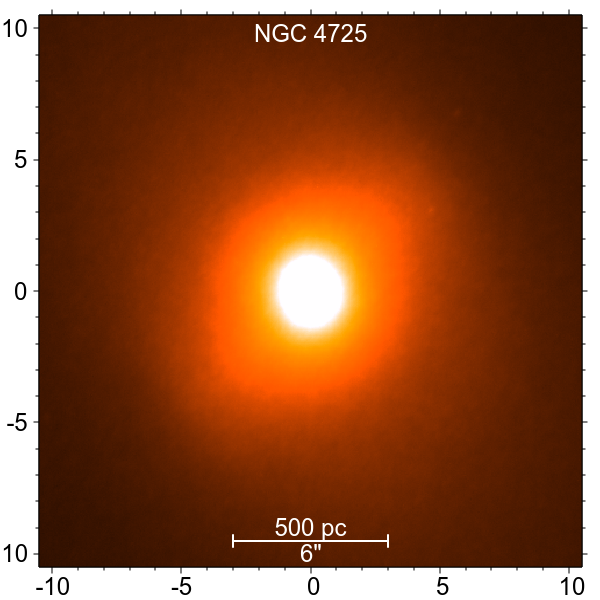}
\includegraphics[width=0.24\textwidth]{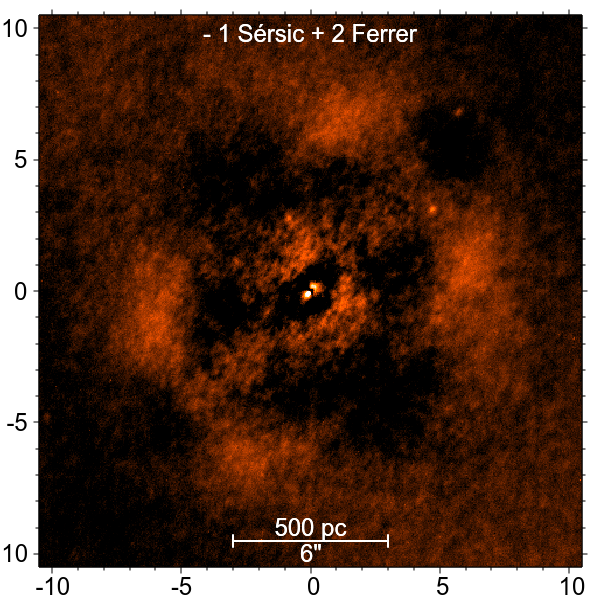}
\includegraphics[width=0.24\textwidth]{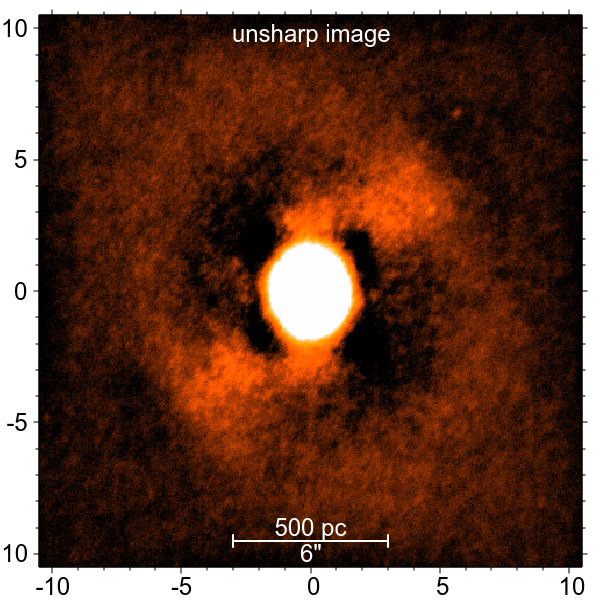}
\includegraphics[width=0.24\textwidth]{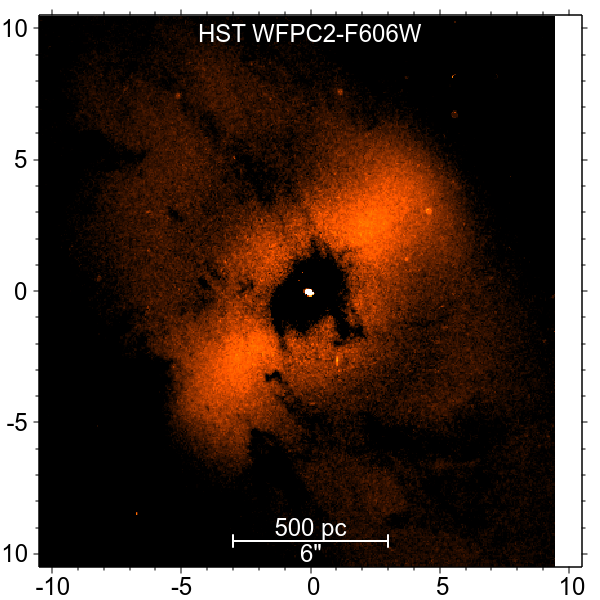}\\
\vspace*{.2cm}
\includegraphics[width=0.24\textwidth]{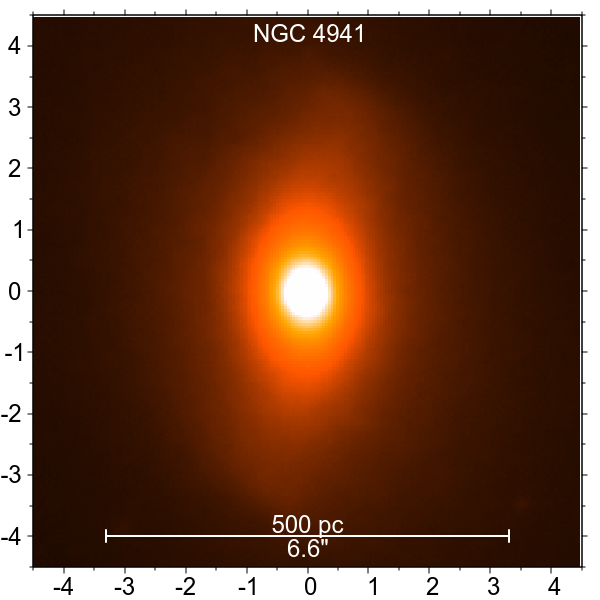}
\includegraphics[width=0.24\textwidth]{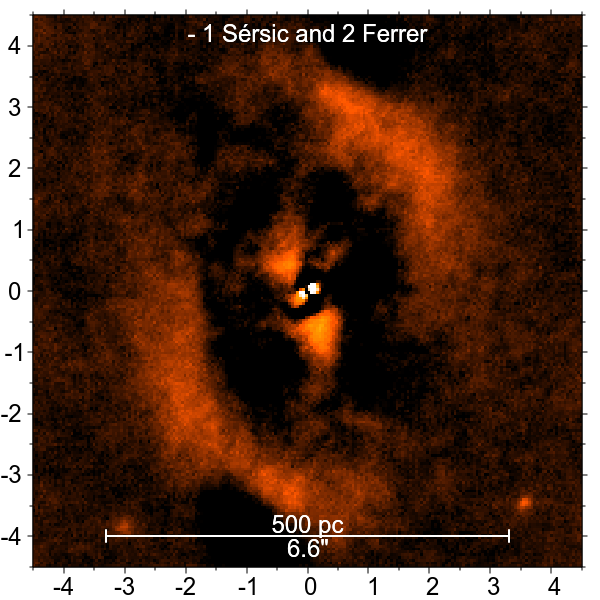}
\includegraphics[width=0.24\textwidth]{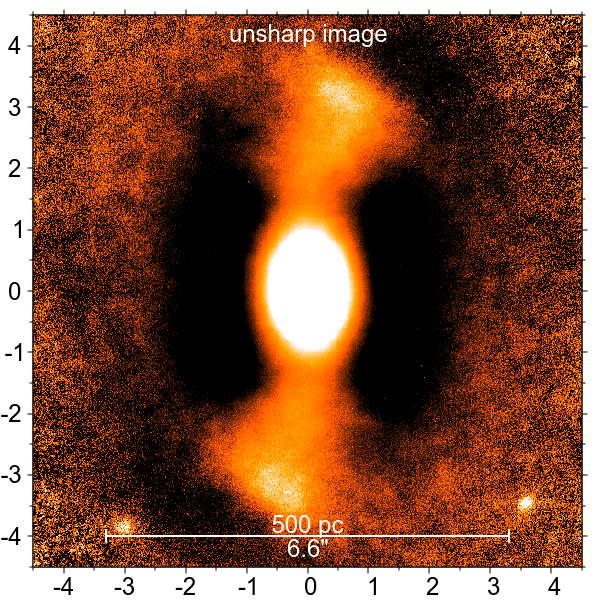}
\includegraphics[width=0.24\textwidth]{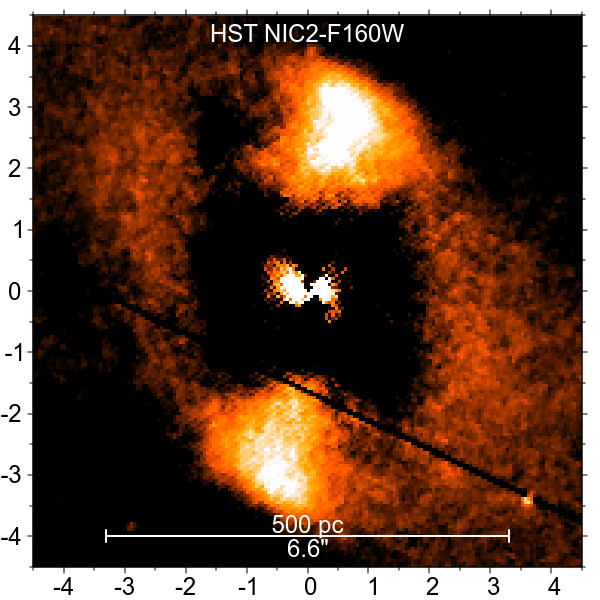}\\
\caption{Same as for Fig. \ref{allfits}, this time for NGC~3486, NGC~3607, NGC~4138, NGC~4725, and for NGC~4941.}
\end{figure*}

\begin{figure*}
 \centering
 \includegraphics[width=0.24\textwidth]{pics2/raltorig5033.png}
\includegraphics[width=0.24\textwidth]{pics2/raltfit5033.png}
\includegraphics[width=0.24\textwidth]{pics2/raltunsh5033.png}
\includegraphics[width=0.24\textwidth]{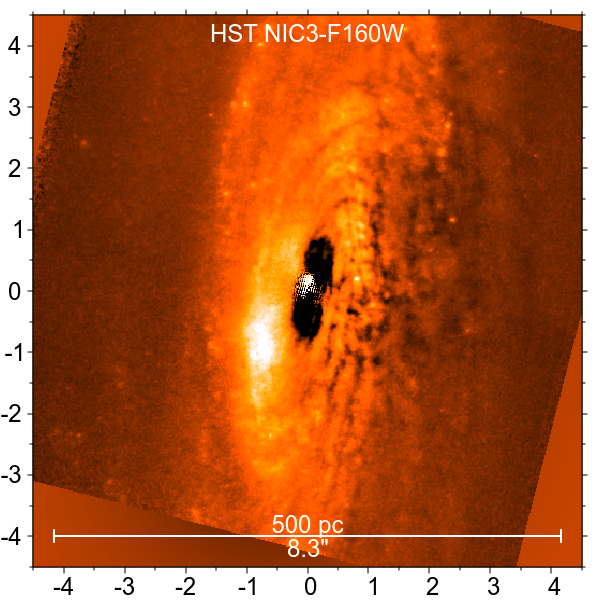}\\
\vspace*{.2cm}
\includegraphics[width=0.24\textwidth]{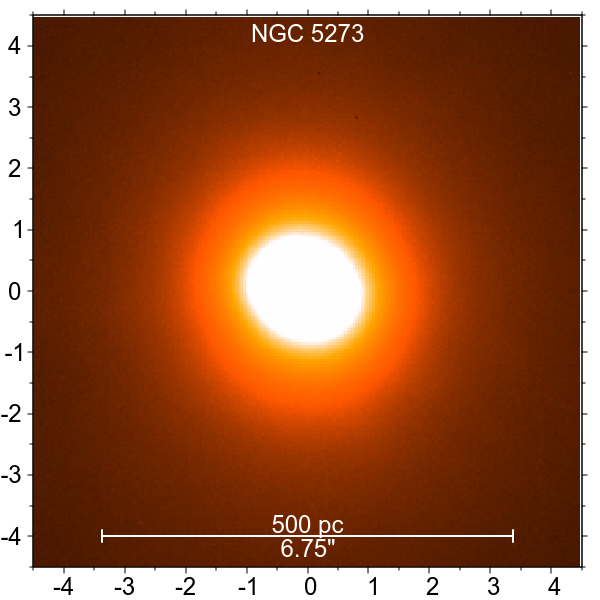}
\includegraphics[width=0.24\textwidth]{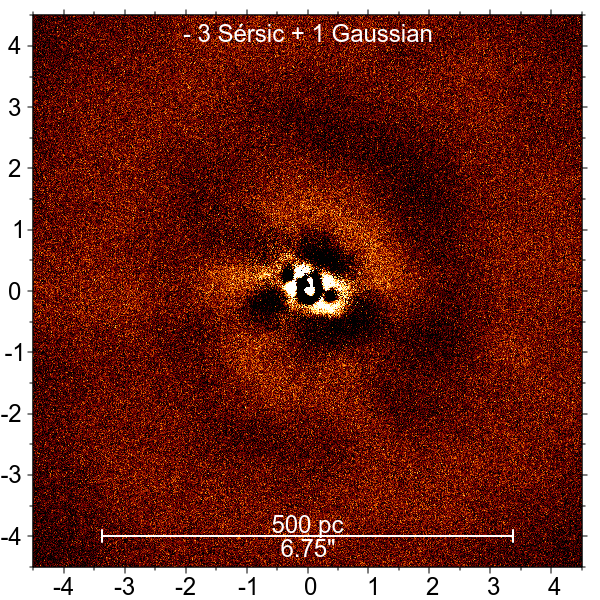}
\includegraphics[width=0.24\textwidth]{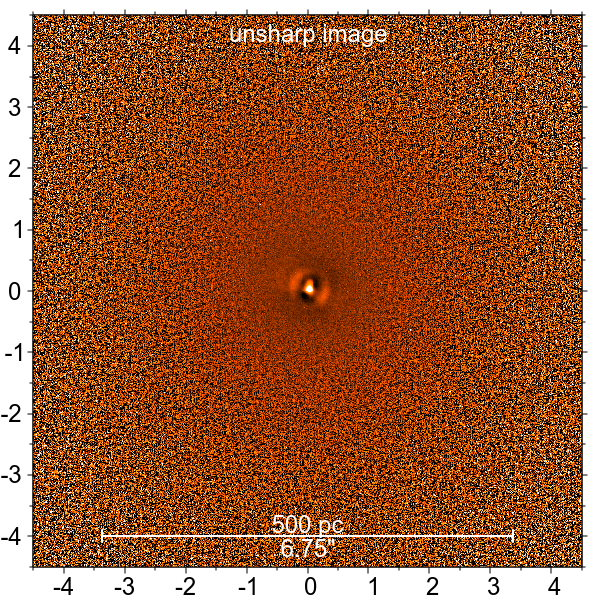}
\includegraphics[width=0.24\textwidth]{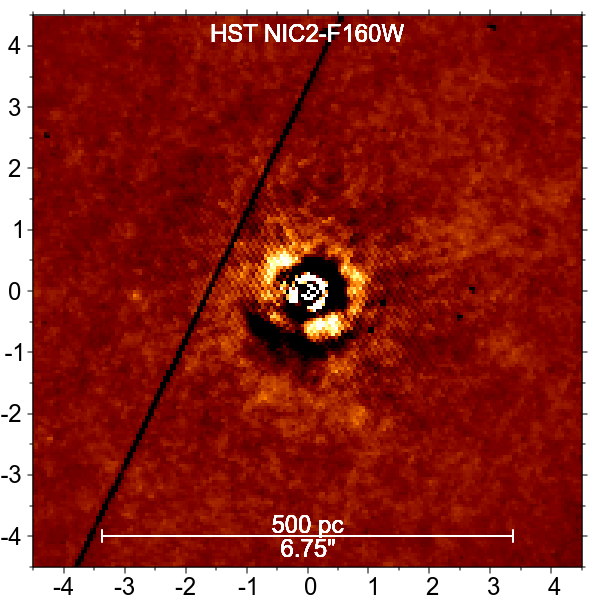}\\
\vspace*{.2cm}
\includegraphics[width=0.24\textwidth]{pics2/raltorig5347.png}
\includegraphics[width=0.24\textwidth]{pics2/raltfit5347.png}
\includegraphics[width=0.24\textwidth]{pics2/raltunsh5347.png}
\includegraphics[width=0.24\textwidth]{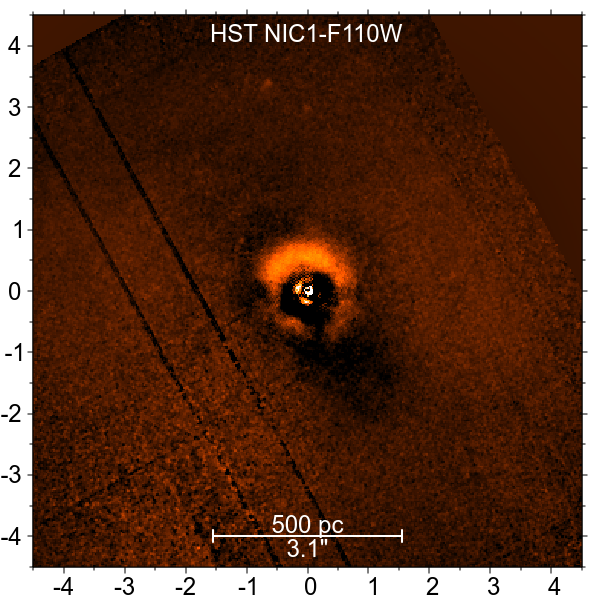}\\
\vspace*{.2cm}
\includegraphics[width=0.24\textwidth]{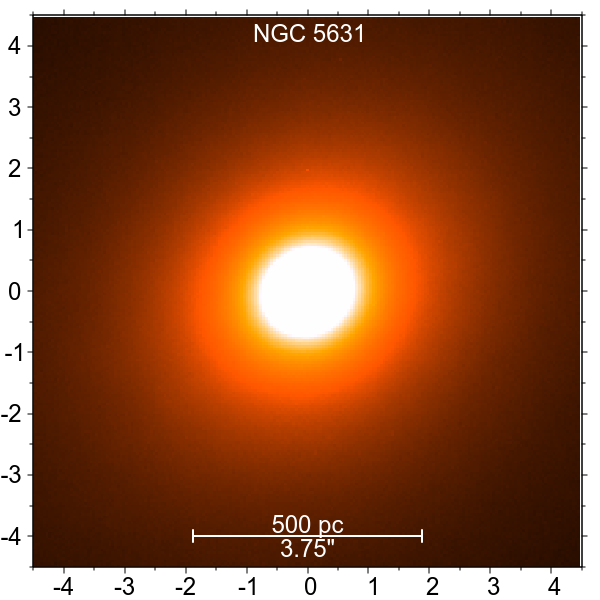}
\includegraphics[width=0.24\textwidth]{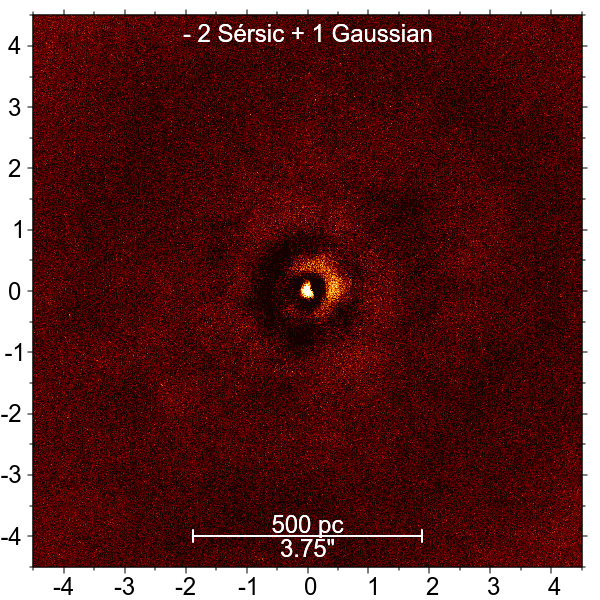}
\includegraphics[width=0.24\textwidth]{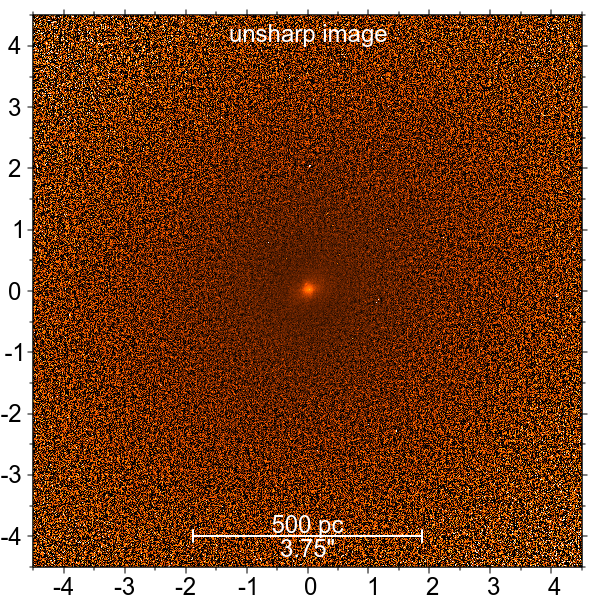}
\includegraphics[width=0.24\textwidth]{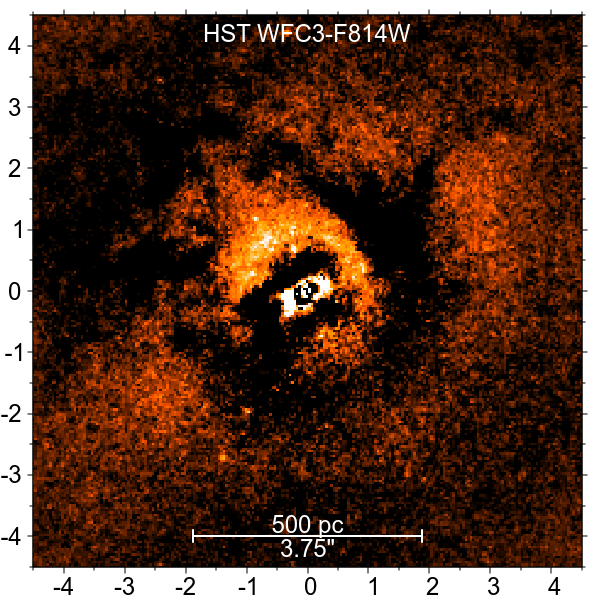}\\
\vspace*{.2cm}
\includegraphics[width=0.24\textwidth]{pics2/raltorig5695.png}
\includegraphics[width=0.24\textwidth]{pics2/raltfit5695.png}
\includegraphics[width=0.24\textwidth]{pics2/raltunsh5695.png}
\includegraphics[width=0.24\textwidth]{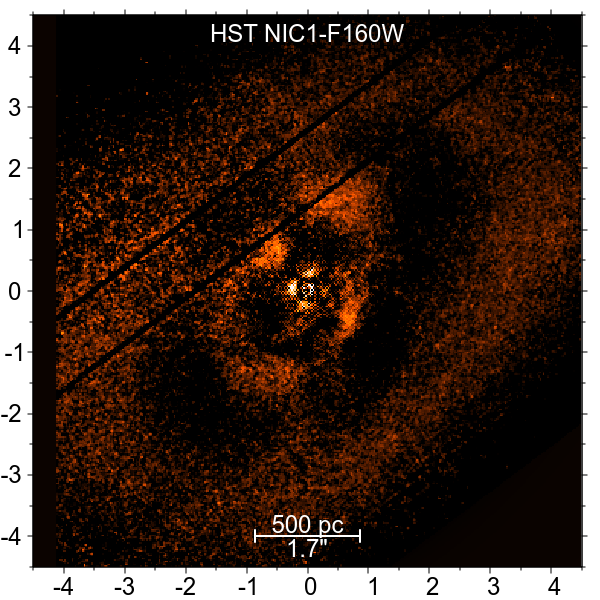}\\
\caption{Same as for Fig. \ref{allfits}, this time for NGC~5033, NGC~5273, NGC~5347, NGC~5631, and for NGC~5695.}
\end{figure*}

\begin{figure*}
 \centering
 \includegraphics[width=0.24\textwidth]{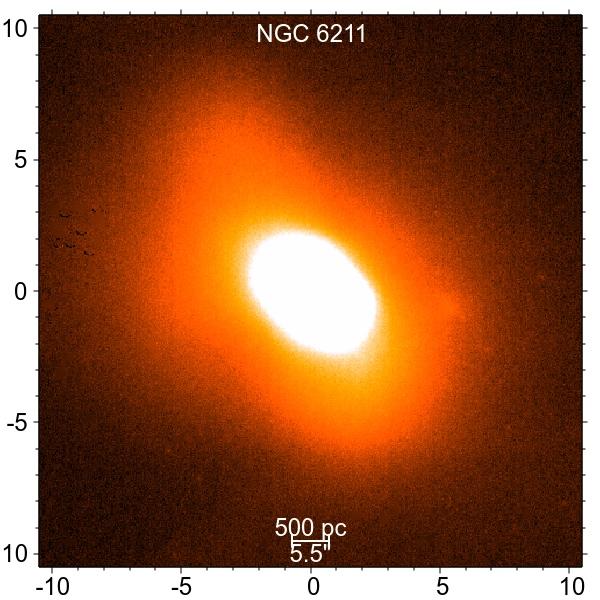}
\includegraphics[width=0.24\textwidth]{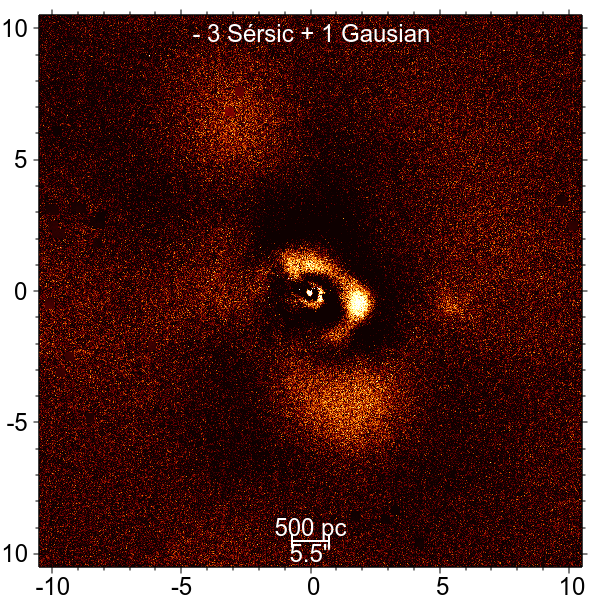}
\includegraphics[width=0.24\textwidth]{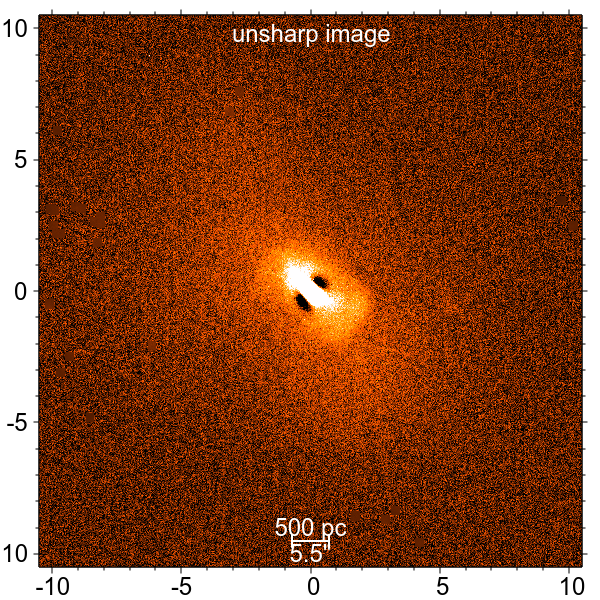}
\includegraphics[width=0.24\textwidth]{pics2/noFit.png}\\
\vspace*{.2cm}
\includegraphics[width=0.24\textwidth]{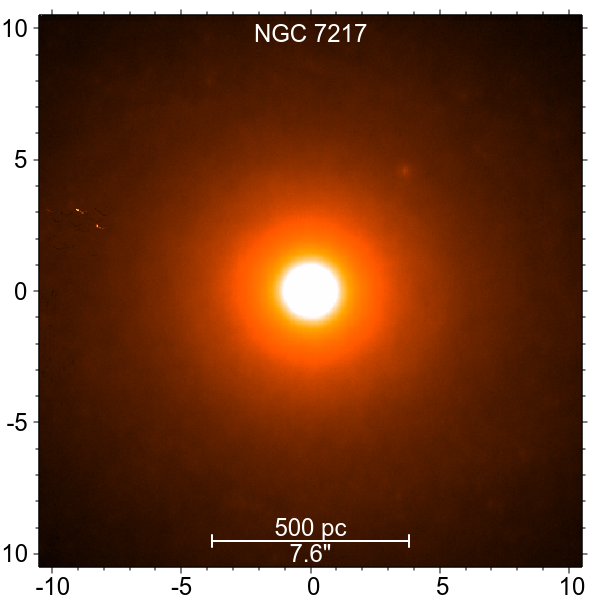}
\includegraphics[width=0.24\textwidth]{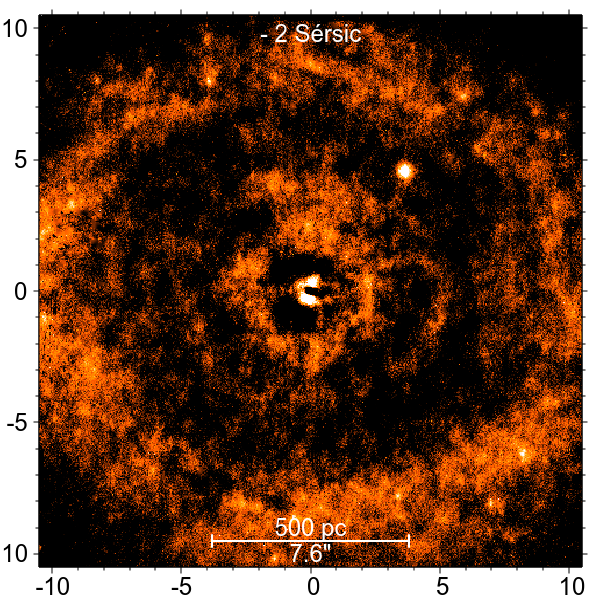}
\includegraphics[width=0.24\textwidth]{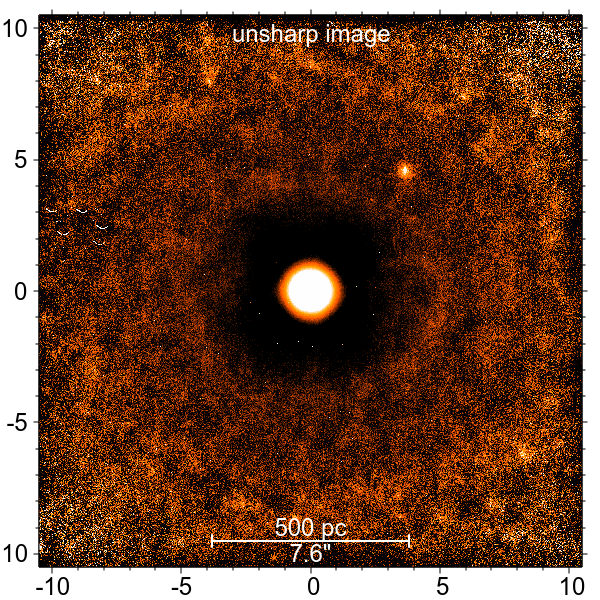}
\includegraphics[width=0.24\textwidth]{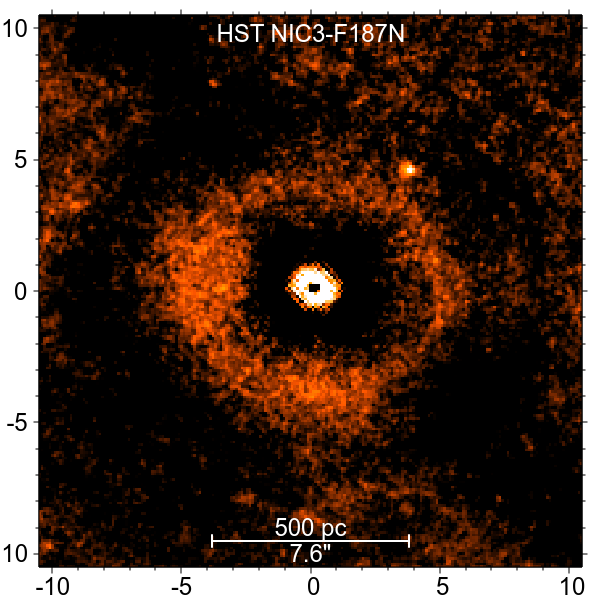}\\
\vspace*{.2cm}
\includegraphics[width=0.24\textwidth]{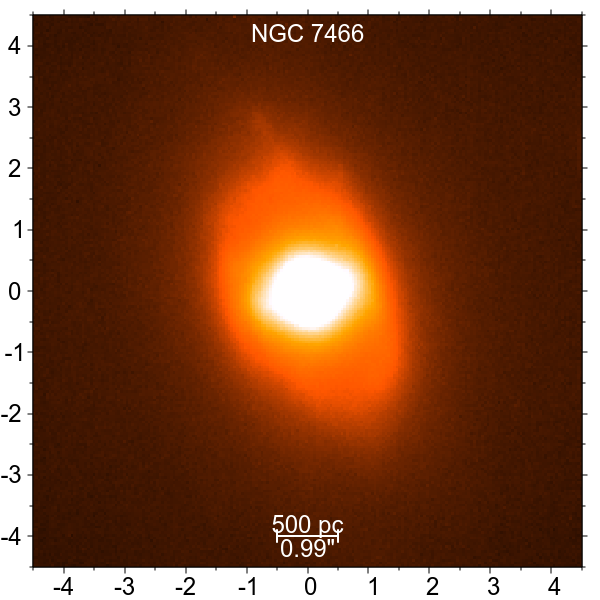}
\includegraphics[width=0.24\textwidth]{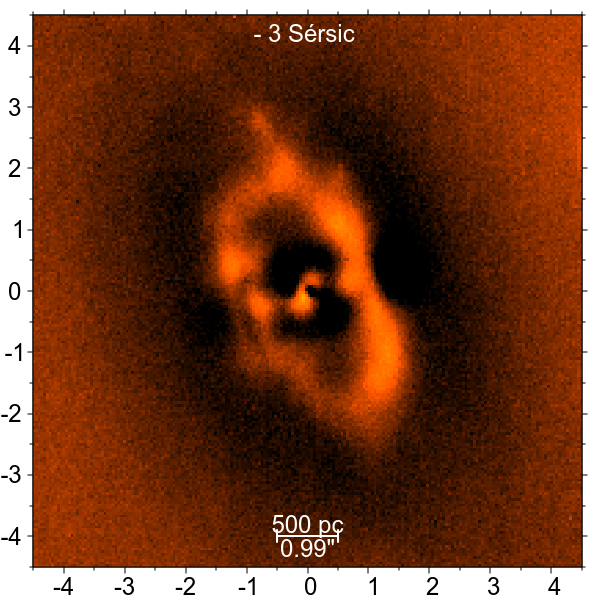}
\includegraphics[width=0.24\textwidth]{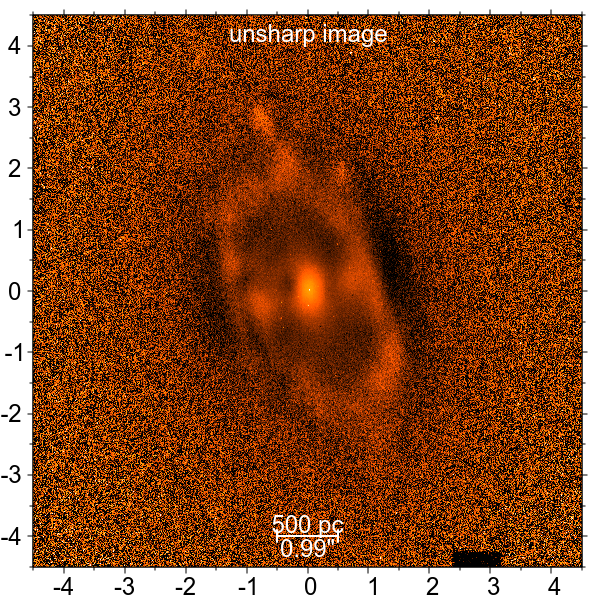}
\includegraphics[width=0.24\textwidth]{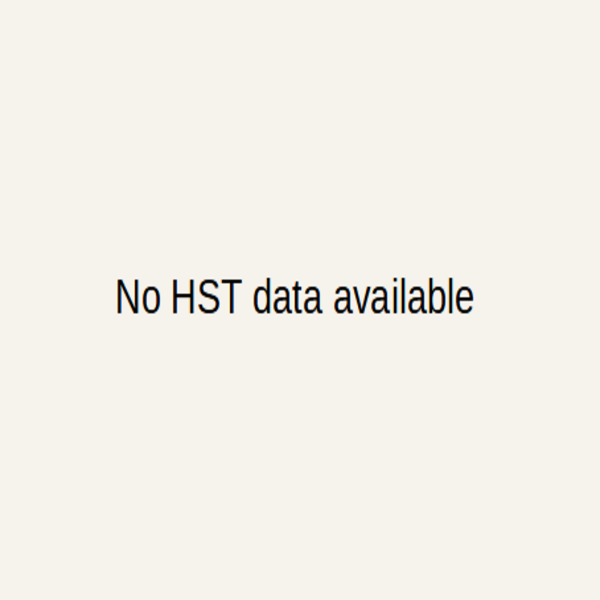}\\
\vspace*{.2cm}
\includegraphics[width=0.24\textwidth]{pics2/raltorig7674.png}
\includegraphics[width=0.24\textwidth]{pics2/raltfit7674.png}
\includegraphics[width=0.24\textwidth]{pics2/raltunsh7674.png}
\includegraphics[width=0.24\textwidth]{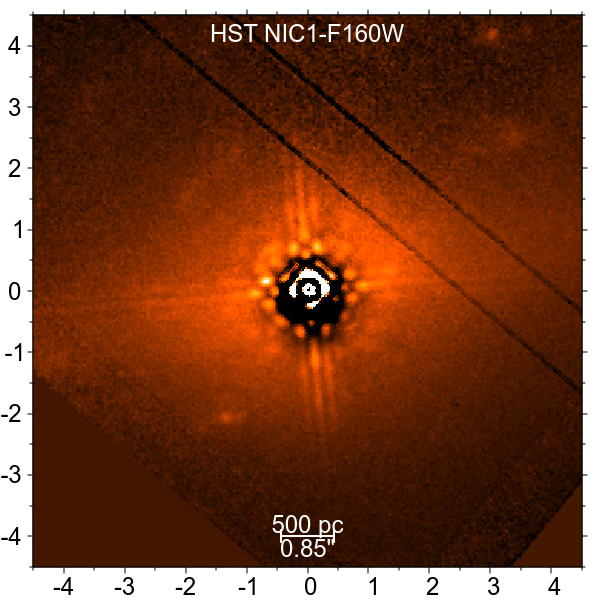}\\
\caption{Same as for Fig. \ref{allfits}, this time for NGC~6211, NGC~7217, NGC~7466, and for NGC~7674.}
\end{figure*}

\newpage
\twocolumn

\section{Notes on individual sources}\label{appendix2}

\begin{figure}[h]
 \centering
\includegraphics[width=0.24\textwidth]{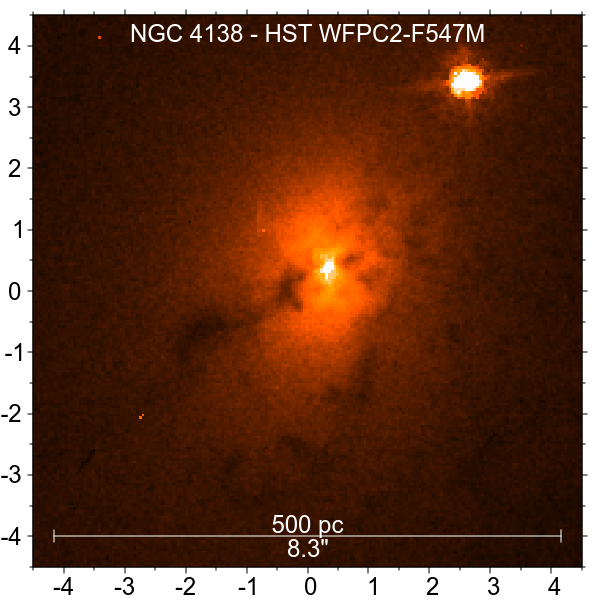}
\includegraphics[width=0.24\textwidth]{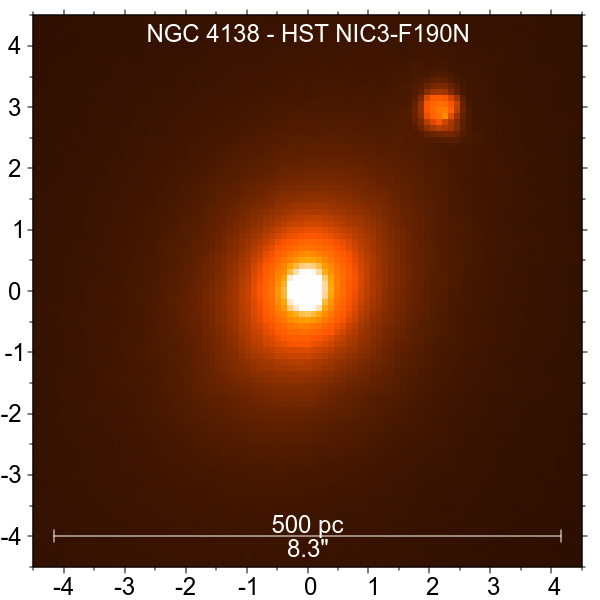}

\caption{HST optical (GO-proposal 6837) and NIR image (GO-proposal 11080) of NGC~4138, respectively. In the optical image, NGC~4138 appears disturbed with two
prominent dust plumes, while it looks fairly smooth in the NIR-image.
North is up and east to the left in both images.}
\label{hst4138}
\end{figure}

In the following, we briefly summarize some characteristics of the sources, describe the results of our fits, and compare them with those reported in previous studies, mostly based on HST images.
\paragraph{MRK~461:} The classification as Seyfert~2 in \cite{2010A&A...518A..10V} is based on \cite{1992ApJ...393...90H}.
It was described as composite (H\,{\sc ii} regions and Seyfert~2 nucleus) in \cite{1999A&AS..135..437G} and \cite{2020ApJ...901..161K}. \\
Best results were obtained after fitting two \ser profiles on our 21" image. A ring-like structure emerges with a radius of \(\sim 5\farcs75\) (\(1.9\,\mathrm{kpc}\)).
 What appears to be a ring-like structure in our images appears rather to be a tightly wound nuclear spiral-like structure in HST UV, optical and NIR images in \cite{1999AJ....118.2646M},
\cite{2002ApJ...569..624P} and \cite{2007AJ....134..648M}, respectively. These spiral-like structures resemble two individual 
spiral arms, which open up at about 5" north and south from the center.
Spiral structure is also indicated in the boxy, twisted isophotes detected in NIR HST-images by \cite{2004ApJ...616..707H}.  
The tight, wound structure is also present in the unsharp mask image, but it is faint. 
\paragraph{NGC~2985:} \cite{2010A&A...518A..10V} listed this galaxy as Sy 1.9, while \cite{1997ApJS..112..315H}
classified NGC~2985 as LINER. Its black hole mass was estimated by \cite{2006AJ....131.1236D} as \(10^{8.2}\,M_\odot\).
Both \cite{2002ApJ...567...97L} and \cite{2008A&A...478..403C} did not detect a bar in the system, while \cite{2010MNRAS.402.2462C} 
may have found a large-scale bar with a length of \(\sim 8.1\,\mathrm{kpc}\).
\cite{2008A&A...478..403C} and \cite{2010MNRAS.402.2462C} discussed the presence of an ultra-compact nuclear ring 
with a semi-major axis \(a = 0\farcs5\) (\(42\,\mathrm{pc}\)) and an ellipticity \(\epsilon = 0.18\) dominated by a young stellar population.  NGC~2985 has two nearby neighbors (NGC~3027 and KDG 59) which could have caused a deviation from axisymmetry in the center of
NGC~2985 and triggered the ultra-compact nuclear star-forming ring \citep{2008A&A...478..403C}.\\
In both - the LBT and HST images - we see an unusually wide ring after subtraction of one \ser profile. However, the ring vanishes completely 
if we add another \ser and one Gaussian profile. The unsharp mask image is featureless, thereby validating the non-existence of any wide ring. Given the predominantly 
young age of the stellar population in the 0\farcs5 nuclear ring and the fact that \cite{2008A&A...478..403C} did not
detect the ring in a \textit{F814W} HST image, it is no surprise 
that we were not able to detect this feature in our even redder bands. Interestingly, we clearly see circularly arranged plumes of dust 
in both the HST \textit{H}-band and the LBT \textit{Ks}-band images, which are more prominent in the \textit{H}-band image.
\paragraph{NGC~3185:} For this barred galaxy, two SMBH estimates exist. \cite{2002ApJ...579..530W} derived
\(10^{6.06}\,M_\odot\) based on stellar velocity dispersion measurements, while \cite{2006AJ....131.1236D}
found \(10^{6.9}\,M_\odot\) using the \textit{K}-band bulge luminosity of NGC~3185.
Based on archival HST data, a star-forming nuclear ring with a semi-major axis of \(a = 2\farcs2\) (\(180\,\mathrm{pc}\)) and an ellipticity of \(\epsilon = 0.37\)
was detected by \cite{2008JPhCS.131a2046C} and discussed in \cite{2010MNRAS.402.2462C}.
Interestingly there seems to be a spatial coincidence of radio emission with the nuclear ring \citep{2019MNRAS.485.3185C}.\\
The subtraction of a single \ser profile was sufficient to clearly reveal the starforming ring.
Modeling with GALFIT we found a semi-major axis \(a = 2\farcs15\) (\(185\,\mathrm{pc}\)) and an ellipticity $\epsilon$ = 0.35. This is basically identical to what has 
been described in
\cite{2010MNRAS.402.2462C}. The thickness of the ring is \(1\farcs65\) (\(135\,\mathrm{pc}\)) and the integrated brightness of the ring 
is \(m_{\mathit{K}} = 13.86\). 
Judging from the HST NIR image, there seems to be a bar-like 
connection between the very center and the ring at \(\mathrm{PA} \sim 45\degr\), but that might be residuals from the imperfect fit. 
Our unsharp mask image confirms the ring too.
\paragraph{NGC~3254:} 
This source has a narrow dust lane close to the nucleus \citep{2004AJ....128.1124S} but was described as featureless in its center otherwise. The SMBH mass has been determined by \cite{2006AJ....131.1236D} to \(10^{7.2}\,M_\odot\).\\
Our image clearly shows a bar-like structure or at least extremely boxy isophotes in the center extending for about 
\(20''\) (\(1.9\,\mathrm{kpc}\)), with a spiral structure 
starting from the ends of the bar-like structure. A simple \ser fit with a high ellipticity resulted in residuals expected for a bar-like structure: overfitting along the major axis and underfitting along the minor axis results in a butterfly-like structure 
(Fig.~\ref{fitexample}). 
The bar-like structure can also easily be seen on the
HST \textit{H}-band images of NGC~3254 taken for studying Cepheids \citep{2022ApJ...934L...7R}. 
A central bar has also been found in the analysis of deep {\it Spitzer} data by \cite{2015ApJS..219....4S}, which derived a bar-length of 26\farcs2. 
Due to the high inclination of NGC~3254 (\(> 60\degr\)),
the bar is almost seen edge-on, resulting in a X-shaped appearance of the central part \citep{2017A&A...598A..10L}.\footnote{\url{https://esahubble.org/images/potw2124a/}} Because of its high inclination and a substantial amount of dust in the center, the presence of a bar may have 
escaped detection in most studies of this source. Our unsharp mask image clearly shows the bar too. Our estimated length of the bar is
about 19", which is similar to what has been derived by \cite{2015ApJS..219....4S}.
\paragraph{NGC~3486:} The nucleus was found to be featureless on HST NIR images \citep{2004ApJ...616..707H}. 
The SMBH mass was determined to be \(10^{6.14}\,M_\odot\) by \cite{2006A&A...455..173P} using stellar velocity dispersion measurements
and to be \(10^{6.5}\,M_\odot\) by \cite{2006AJ....131.1236D} via the \textit{K}-bulge luminosity, respectively.
The fitting was hampered by the nearby galaxy about 4\farcs4 NE of NGC~3486, which was masked out.
Fitting one \ser showed the same butterfly structure as in NGC~3254, indicating the presence of a bar.
\cite{2013ApJ...771...59M} found a bar with a length of about 25" at a PA \(\sim 75\degr\) from their 1D profile analysis, 
which is consistent with our observations, despite the smaller FoV.
\cite{2015ApJS..219....4S} did not detect one from the same data set in their 2D analysis with GALFIT. Also \cite{2021MNRAS.507.5952R} did not require a bar-like component for their fits on NGC~3486 using 2MASS data. Our unsharp mask image does not reveal a bar most likely 
because it extends beyond the size of our image.
The HST \textit{F160W} image shows qualitatively the same morphology as our LBT image.
\paragraph{NGC~3607:} This galaxy is the central and brightest member of the Leo II group
\citep[and references therein]{2007A&AT...26..311A},
forms a physical pair with NGC~3608 and is a member of a compact galaxy group within the Leo II group \citep{2015MNRAS.446..120D}. Its classification as a Seyfert~2 galaxy has been questioned by \cite{1997iagn.book.....P}. It could be that the strong low-ionization lines in this source arise from cooling flows, starburst-driven winds and shock-heated gas which would favor the classification as LINER \citep{1987IAUS..121..421H}.  \cite{2009ApJ...695.1577G} derived 1.2 \(\times 10^8\,M_\odot\) for the SMBH of NGC~3607. \cite{2005AJ....129.2138L}
stated that a central dust ring is ‘starting to be organized’ in this source.\\
Even in the \textit{K} band, the dust across the nucleus seriously affected the fits and is clearly visible already on the original image. The best result was obtained after fitting 3 \ser profiles to the nucleus. The residual image resembles pretty 
much what can be seen in the center of the optical HST image in \cite{2005AJ....129.2138L} (their Fig.~20). \\
The ring-like structure, which could also be a tightly wound, almost closed spiral arm, has a semi-major axis of \(\sim 5\farcs8\) (\(370\,\mathrm{pc}\))
and $\epsilon$ = 0.3. Our unsharp mask image shows qualitatively the same behavior. A radius of $\sim$ 5\farcs8 roughly matches
the size of the kinematically decoupled core in NGC~3607 detected by \cite{2007A&AT...26..311A}.
\paragraph{NGC~4138:} 
This galaxy has a ring with a semi-major axis \(a = 19\farcs5\) \citep{1980AJ.....85..637D}.
In addition, it has a counter-rotating disk at \(R \sim 22''\) whose peak intensity coincides with
the peak of the HI distribution and a ring of H\,{\sc ii} regions \citep{1996AJ....112..438J}.
It may also have a nuclear ring of ionized gas at \(R = 4''\) (\(245\,\mathrm{pc}\)) housing a compact stellar minibar \citep{2002AJ....124..706A}. This is one of the few cases, where an unbarred galaxy hosts a nuclear ring. \cite{2006A&A...455..173P} found \(10^{7.75}\,M_\odot\) for the mass of the SMBH of NGC~4138.\\
The analysis was hampered by a star about \(3\farcs6\) NW of the center. Using this star we measured a FWHM of \(\sim 0\farcs079\), i.e., we were not far from being diffraction-limited on that image. The galaxy appears lenticular and undisturbed in the center.
In spite of the excellent image quality, we could not derive a good fit. 
Our ‘best’ fit was obtained with one \sser, leaving residuals indicative of a second component (underfitting in the central arcsec, underfitting up to a radius of \(4''\)). Whatever we tried, GALFIT was not able to fit two components simultaneously,
it always crashed. Our badly fitted region with a radius of about \(4''\) may be the exponential bulge component with 
a scale length of \(r_0 = 1\farcs9\) \citep{1996AJ....112..438J}.  A medium-band HST \textit{F547M}-image  shows a completely different view. The galaxy looks very disturbed in the center with at least two plumes of dust coming out of
(or going into) it (see Fig.~\ref{hst4138}). It is not clear, whether the disturbed appearance is due to the large amounts of dust in the center or indicative of a merger remnant as was speculated by \cite{2002AJ....124..706A} based on their findings of a chemically
distinct nucleus in NGC~4138. Our unsharp mask image is featureless.

\paragraph{NGC~4725:} This galaxy is tidally interacting with the less massive galaxy  NGC~4747 \citep{1984A&A...140..125W} and is a member of the optical galaxy triplet HOLM 468 \citep{1937AnLun...6....1H}.
It hosts at least two bars. One at larger scales with a semi-major axis of \(a = 118''\) (\(9.9\,\mathrm{kpc}\)) and an inner one with a semi-major axis of \(a = 5\farcs6\) (\(470\,\mathrm{pc}\)) at \(\mathrm{PA} = 141\degr\), lying within a kinematically confirmed inner disk
\citep{2004A&A...415..941E,2005MNRAS.364..283E}. This galaxy hosts an inner ring with a semi-major axis of \(a = 2\farcm18\) (\(10.8\,\mathrm{kpc}\); \citealt{1988ApJS...66..233B}). Its SMBH mass was determined to be \(10^{7.49}\,M_\odot\) by \citep{2003MNRAS.345.1057M}.\\
The inner bar discussed by \cite{2004A&A...415..941E} can already be spotted on our original \textit{Ks}-band image and is clearly detected 
after subtraction of one \ser profile. The additional subtraction of two Ferrer profiles shows
a faint nuclear ring with a radius of \(6\farcs7\) (\(555\,\mathrm{pc}\)) that encircles the inner bar. 
The ring is also qualitatively present in our unsharp and the HST \textit{V}-band image.
\paragraph{NGC~4941:} This ringed galaxy has a SMBH with a mass of \(10^{6.91}\,M_\odot\) \citep{2011A&A...536A..36A}.
An inner bar with a radius of \(3\farcs5\) (\(110\,\mathrm{pc}\)) has been detected by \cite{2002ApJ...567...97L}, which was also seen by 
\cite{2000A&AS..145..425G}, \cite{2004ApJ...616..707H} and discussed in \cite{2024MNRAS.528.3613E}. It may also have a large-scale 
bar \citep{2004A&A...415..941E}.\\
The inner bar can also be glimpsed on our original \textit{Ks}-band image with similar properties as described by \cite{2024MNRAS.528.3613E}.
Subtraction of \ser profiles already indicates a nuclear ring in this system encircling the inner bar. This ring becomes more prominent after subtracting
one \ser and two Ferrer profiles, the latter of which were used to describe the inner bar. The ring has a radius of about 
\(2\farcs8\) (\(210\,\mathrm{pc}\)). The bar dominates in the unsharp mask image. The HST image clearly shows the ring too.
\paragraph{NGC~5033:} This is a highly inclined galaxy, whose Sy-type ranges from 1.2 to 1.9 \citep{1993ApJ...414..552O,1997ApJS..112..315H,2010ApJ...725.1749T}. Its SMBH mass is 
\(10^{7.3}\,M_\odot\) \citep{2003MNRAS.345.1057M}. After examination of optical HST images
\cite{2002ApJ...567...97L} suggested that NGC~5033 hosts a triple bar system. On the contrary, \cite{2001ApJ...562..139M} did not find any bar in their analysis of NIR HST data, while the bar strengths  
determined by \cite{2002MNRAS.331..880L} on 2MASS images are borderline between a ‘no bar’ or ‘weak bar’ interpretation.
The problem with the analysis is clearly the high inclination and the very strong spiral dust lanes  of this galaxy \citep{2003ApJS..146..353M}. \cite{2004ApJ...616..707H} found isophote twists and boxy isophotes as well as indications for dusty filaments  in their HST \textit{H}-band image.\\
Even our \textit{Ks}-band image shows dusty filaments towards the very center. This and the high inclination hamper performing a good fit. The residual image after fitting 3 \ser profiles shows the spiral structure in the central \(20''\) as seen 
in optical images. On top of this we see a twisted structure connecting the spiral structure with the very center.
If real, it could be an inner spiral or small bar through which material for feeding the SMBH is delivered.
\paragraph{NGC~5273:} \cite{1993ApJ...414..552O} classified NGC~5273 as Seyfert 1.9, while \cite{2010ApJ...725.1749T} derived 
Seyfert type 1.5 after correcting for the host galaxy flux. \cite{2014ApJ...796....8B} measured a black hole mass of 
\(10^{6.67}\,M_\odot\) and \cite{2023ApJ...949...13M} derived \(10^{6.7\text{--}7.3}\,M_\odot\). Little is known about the nuclear morphology. \cite{2010MNRAS.405.1089L} found a weak nuclear bar, but could/did not fit it. \cite{2004ApJ...616..707H} found an isophote twist in this system, which could be due to nested bars or misaligned bars at different pattern speeds.\\
This was one of the most challenging objects to fit since several components were required, but GALFIT did not allow the centers of the components to remain fixed. The latter resulted in considerable substructure even after subtraction of three \ser and one Gaussian profiles (LBT) or three \ser profiles (HST). On top of that, the unsharp mask image shows a different morphology, namely a very compact ring-like structure with a radius of  \(0\farcs27\) (\(20\,\mathrm{pc}\)). Since the results are so ambiguous we are sceptical that any of the substructures on the 
residual images is real.
\paragraph{NGC~5347:}  This barred Seyfert~2 galaxy hosts a fairly massive SMBH of \(10^{8.73}\,M_\odot\) \citep{2016ApJ...827...81I}. 
Interestingly, \cite{2017MNRAS.470...20S} detected a kinematically cold nuclear disk in NGC~5347.  \cite{2003ApJS..146..353M} 
recognized two, symmetric dust lanes forming a very well defined grand-design nuclear dust spiral. 
Evidence of nuclear dust structure was also found by \cite{2004ApJ...616..707H}.\\
Our \textit{Ks}-band image did not show any obvious evidence of dust.
The subtraction of two \ser profiles shows an almost perfectly closed \(m_{\mathit{K}} = 17.0\) ring with a radius of about \(1''\) (\(160\,\mathrm{pc}\)) and a thickness of \(0\farcs5\) (\(80\,\mathrm{pc}\)).  Fitting a single \ser profile to the HST NIC1 \textit{H}-band image displays a ring-like structure as residual as well, though not closed. The latter may be due to the nuclear dust structure found in the studies listed above. The ring is also present in our unsharp mask image.
Redoing the fits by adding a Gaussian profile or substituting one \ser by a Gaussian profile did not change the outcome. Thus, the detection of the ring is robust.
\paragraph{NGC~5631:} A dynamically interesting system, with a flat, large-scale disk, an inclined, inner counter-rotating gaseous disk on \(10\text{--}30''\) (\(300\text{--}900\,\mathrm{pc}\)) scales and an \(8''\) (\(220\,\mathrm{pc}\)) bulge \citep{2009ApJ...694.1550S}. The counter-rotating disk may have formed through a minor merger with a gas-rich satellite.\\
This galaxy exhibits possibly a very weak ring after subtraction of two \ser profiles, which disappears mostly when a Gaussian profile is included into the fit. The unsharp mask image appears featureless, supporting the absence of a stellar ring in the center.
The residuals after the best-fit to the HST \textit{F814W}-image, including two \ser and one Gaussian profile, show quite some residuals, including potential plumes of dust. 
This may be indicative of a potential minor merger event in NGC~5631 discussed by 
\cite{2009ApJ...694.1550S}.
\paragraph{NGC~5695 = MRK~686:} The redshift of \(z = 0.0125\) listed in the QSO-catalog of \cite{2010A&A...518A..10V} is from \cite{1993ApJ...414..552O}. 
More recently \cite{2016A&A...595A.118V} determined \(z = 0.0142\) from 21 cm observations. The SMBH mass in this barred galaxy has been estimated to \(10^{7.7}\,M_\odot\) \citep{2022MNRAS.510.5102O}. The semi-major axes of the bar is \(a = 11''\) (\(3\,\mathrm{kpc}\); \citealt{2002ApJ...567...97L}).
\cite{1998ApJS..117...25M} detected a nuclear dust lane towards the southwest and an inclined spiral in the optical, 
\cite{2004ApJ...616..707H} found an isophote twist in the center in the NIR, but did not comment on it further. \\
We do not see any evidence of dust in our \textit{Ks}-band image, but have a similar isophote twist as \cite{2004ApJ...616..707H}.
Our fits after subtraction of two \ser profiles show a highly elliptical,  
ring-like structure with spiral arms emerging from the ends of the major axis moving outwards.
The ends of the minor axis of the structure 
seem to be connected to the very center via inner spirals. The same can be seen on the HST \textit{F160W} image and also qualitatively in our unsharp mask image. The elliptical ring-like structure has a major axis of \(4\farcs2\) (\(1.22\,\mathrm{kpc}\)) and \(\epsilon = 0.65\).
\paragraph{NGC~6211:} Very little is known about this galaxy. It forms a close pair with NGC~6213
\citep{2002AJ....123.2280P} and has apparently a faint tidal tail \citep{2022MNRAS.510.4608K}.
\cite{1998ApJS..117...25M} noted a nuclear shell indicative of a very late stage of a merger.\\
Our inspection of the  HST \textit{F606W} image from \cite{1998ApJS..117...25M} clearly shows a disturbed banana-shaped morphology which is also present
in our \textit{Ks}-band image. The residuals of our best-fit show considerable substructure, possibly a merger remnant as indicated by a tidal tail.
Three \ser and one Gaussian components were necessary to describe the underlying smooth light distribution. 
The unsharp mask image does not show the unusual structure that clearly appears in the other images.
We were not able to fit any \ser profiles to the HST image.
\paragraph{NGC~7217:} It is not clear whether this source is a LINER or a Seyfert~2. \cite{1997ApJS..112..315H}, for example, classified it as LINER 2 (meaning that it could also be a Seyfert~2), while it is listed as ‘Seyfert 3 or LINER’ in \cite{2010A&A...518A..10V}.
Its SMBH mass was estimated to be \(10^{7.48}\,M_\odot\) by \cite{2007A&A...464..553K}. Because of its morphological characteristics - 
two exponential disks, a counter-rotating stellar disk and at least 7 rings of various types (stellar, CO, H$\alpha$ or dust) with diameters between \(10''\) and \(70''\) and some nuclear spirals - it is a prime target to study fueling in AGNs \citep{2004A&A...414..857C,2006AJ....131.1336S}. \\
We detected two rings in our residual HST and LBT images, each after subtraction of two \ser components. One with a radius of roughly \(10''\) and one with a radius of \(\sim 2\farcs9\) (LBT) and \(4\farcs5\) (HST). It is not fully clear if the ring at \(10''\) in our image coincides with the nuclear H\(\alpha\)/stellar ring with \(r = 10\farcs7\) detected by \cite{1989ApJS...71..433P} and \cite{1995ApJ...450..593B} or the dust ring with \(r = 8\farcs6\) discussed in
\cite{1995ApJ...450..593B}. Since the  LBT \(\sim 2\farcs9\) and HST \(4\farcs5\) (HST) rings have different diameters and do not appear in the unsharp mask image, they are unlikely to be real. We also do not claim to have detected either the
H\(\alpha\)/stellar ring with \(r = 10\farcs7\) detected by \cite{1989ApJS...71..433P} and \cite{1995ApJ...450..593B} or the dust ring with \(r = 8\farcs6\) discussed in \cite{1995ApJ...450..593B}, since a ring at these radii does not show up on both - 
the GALFIT-processed residual and the unsharp images.
\paragraph{NGC~7466 = MRK~1127:} This unbarred galaxy is the only one in our sample for which no HST data or other high-resolution data exist.\\
This galaxy has a highly elliptical nuclear ring, which can already be spotted on the original image. This is one of the few cases where we could fit a ring with a 
semi-major axis  of \(2\farcs6\) (\(1.31\,\mathrm{kpc}\)), an ellipticity of \(\epsilon = 0.6\), and an integrated brightness of \(m_{\mathit{K}} = 13.53\) once three \ser
profiles had been subtracted.
 The ring shows some clumpy substructure, which is also clearly present in the unsharp mask image. Galaxies without nonaxisymmetric features hosting a nuclear ring as the one here are not uncommon \citep{2010MNRAS.402.2462C}. If deprojected, the ring appears relatively round.
\paragraph{NGC~7674 = MRK~533:} This barred galaxy is the brightest of the Hickson 96 group, dominated by two large and two considerably smaller members \citep{1997A&A...321..409V}. According to \cite{2017NatAs...1..727K}, this system may host a binary SMBH system, although this has been  
questioned more recently by \cite{2022ApJ...933..143B}. 
It shows filamentary or wisp-like structures, a partial ring, and an inner spiral pattern in optical HST data
\citep{1998ApJS..117...25M,1999AJ....118.2646M,2020ApJ...893...26S}. Disky, twisted isophotes were detected in NIR HST-images by \cite{2004ApJ...616..707H}.  \citet{2015ApJ...812..117A} found a ring of molecular gas in this source.\\
Contrary to the HST \textit{F606W} images, NGC~7674 looks fairly symmetric and undisturbed in its central part in our image. 
The residuals after fitting two \ser components and one Gaussian profile show only minor PSF subtraction artifacts.
This galaxy, with a FWHM of \(0\farcs24\),
has one of the most compact cores of our sample supported by the presence of diffraction spikes. 
An unresolved core supported by the presence of clear diffraction ring has been noted by \cite{2001ApJ...547..129Q} 
in the \textit{F160W} image of NGC~7674. Our inspection of the \textit{F160W} image shows the source to be similarly featureless as our LBT image.
The unsharp mask image is featureless as well.

\end{appendix}

\end{document}